\documentclass{article}

\PassOptionsToPackage{sort, numbers, compress}{natbib}

    \usepackage[final]{neurips_2020}

\usepackage[utf8]{inputenc} 
\usepackage[T1]{fontenc}    
\usepackage{hyperref}       
\usepackage{url}            
\usepackage{booktabs}       
\usepackage{amsfonts}       
\usepackage{nicefrac}       
\usepackage{microtype}      
\usepackage{wrapfig}
\usepackage{amsmath}
\usepackage{bm}
\usepackage{mathtools}
\usepackage[textsize=tiny]{todonotes}
\usepackage{enumitem}

\usepackage{subcaption}

\usepackage{algorithm}
\usepackage{algorithmicx}
\usepackage{algpseudocode}

\usepackage{textcomp}
\usepackage{setspace}

\usepackage{siunitx}
\newcommand{\Ex}{\mathbb{E}}                                                   %
\DeclareMathOperator{\cov}{Cov}

\renewcommand{\d}{\mathrm{d}}
\newcommand{\ddx}{\frac{\d}{\d x}}

\newcommand{\re}{\mathbb{R}}

\DeclareMathOperator*{\argmin}{arg\,min}

\newcommand{\Cc}{\mathcal{C}}
\newcommand{\Dc}{\mathcal{D}}

\newcommand{\Fc}{\mathcal{F}}

\newcommand{\Sc}{\mathcal{S}}

\newcommand{\Tc}{\mathcal{T}}

\newcommand{\xb}{\boldsymbol{x}}
\newcommand{\Yb}{\boldsymbol{Y}}
\newcommand{\yb}{\boldsymbol{y}}

\newcommand{\Zb}{\boldsymbol{Z}}
\newcommand{\zb}{\boldsymbol{z}}

\newcommand{\sbb}{\boldsymbol{s}}

\newcommand{\Lambdab}{\boldsymbol{\Lambda}}

\newcommand{\kappab}{\boldsymbol{\kappa}}
\newcommand{\lambdab}{\boldsymbol{\lambda} } %

\newcommand{\Eb}{\boldsymbol{E}}
\newcommand{\Dkl}{\mathcal{D}_{\rm KL}}         %
\newcommand{\Jc}{{\cal J}}
\newcommand{\Lc}{{\cal L}}

\newcommand{\Nc}{{\cal N}}

\def\zal2{z_{\alpha/2}}

\newcommand{\R}{{\mathbb{R}}}

\DeclareMathOperator{\trace}{Tr}
\DeclareMathOperator{\Span}{span}

\newcommand{\bfF}{\mathbf{F}}

\newcommand{\sfL}{\mathcal{L}}

\newcommand{\sfN}{\mathcal{N}}

\newcommand{\bff}{\mathbf{f}}

\newcommand{\bfh}{\mathbf{h}}

\newcommand{\bfx}{\mathbf{x}}

\usepackage{amsthm} 

\newcommand{\chapternote}[1]{{%
  \let\thempfn\relax
  \footnotetext[0]{#1}
}}

\newtheorem{proposition}{Proposition}

\newcommand\dkl{\mathcal{D}_{\text{KL}}}

\title{Greedy inference with structure-exploiting lazy maps}

\author{%
	  Michael C.~Brennan \thanks{These authors contributed equally to this work.
}\\
	Massachusetts Institute of Technology\\
	Cambridge, MA 02139 USA\\
	\texttt{mcbrenn@mit.edu}\\
	\And
  Daniele Bigoni \footnotemark[1]\\
  Massachusetts Institute of Technology\\
  Cambridge, MA 02139 USA\\
  \texttt{dabi@mit.edu}\\
  \And
  Olivier Zahm\\
  Universit\'{e} Grenoble Alpes, INRIA, CNRS, LJK\\
  38000 Grenoble, France\\
  \texttt{olivier.zahm@inria.fr}\\
  \And
  Alessio Spantini\\
  Massachusetts Institute of Technology\\
  Cambridge, MA 02139 USA\\
  \texttt{alessio.spantini@gmail.com}
  \And
  Youssef Marzouk\\
  Massachusetts Institute of Technology\\
  Cambridge, MA 02139 USA\\
  \texttt{ymarz@mit.edu}
}

\begin{document}

\maketitle

\begin{abstract}
	We propose a framework for solving high-dimensional Bayesian inference problems using \emph{structure-exploiting} low-dimensional transport maps or flows. These maps are confined to a low-dimensional subspace (hence, lazy), and the subspace is identified by minimizing an upper bound on the Kullback--Leibler divergence (hence, structured). Our framework provides a principled way of identifying and exploiting low-dimensional structure in an inference problem. It focuses the expressiveness of a transport map along the directions of most significant discrepancy from the posterior, and can be used to build deep compositions of lazy maps, where low-dimensional projections of the parameters are iteratively transformed to match the posterior. We prove weak convergence of the generated sequence of distributions to the posterior, and we demonstrate the benefits of the framework on challenging inference problems in machine learning and differential equations, using inverse autoregressive flows and polynomial maps as examples of the underlying density estimators.
\end{abstract}


\section{Introduction}
\label{sec:intro}
Inference in the Bayesian setting typically requires the computation of integrals $\int f  \, \d\pi$  over an \textit{intractable} posterior distribution whose density\footnote{In this paper, we only consider distributions that are absolutely continuous with respect to the Lebesgue measure on $\mathbb{R}^d$, and thus will use the notation $\pi$ to denote both the distribution and its associated density.} $\pi$ is known up to a normalizing constant.
One approach to this problem is to construct a
deterministic nonlinear transformation, i.e., a \emph{transport map}
\cite{villani2008optimal},
that induces a coupling of $\pi$ with a tractable distribution
$\rho$ (e.g., a standard Gaussian).
Formally, we seek a map $T$ that pushes forward $\rho$ to $\pi$, written as $T_\sharp \rho = \pi$, such that the change of variables
$\int f \, \d \pi = \int f \circ T \, \d \rho $ makes integration
tractable.

Many constructions for such maps have been developed in recent years. Normalizing flows (see \cite{rezende2015variational,tabak2013family,papamakarios2019normalizing,kobyzev2020normalizing} and references therein) build transport maps via a deep composition of functions parameterized by neural networks, 
with certain ansatzes 
 to enable efficient computation. Many recently proposed autoregressive flows (for example \cite{dinh2016density,Kingma2016,papamakarios2017masked,huang2018neural,de2019block}) compose triangular maps, which allow for efficient evaluation of Jacobian determinants. In general, triangular maps \cite{bogachev2005triangular,rosenblatt1952remarks,knothe1957contributions} are
sufficiently general to couple any absolutely continuous pair of
distributions $(\rho, \pi)$, and their numerical approximations have
been investigated in \cite{el2012bayesian,marzouk2016introduction, spantini2018inference, Jaini2019}. The flow map of a neural ordinary differential equation \cite{Chen2018,Dupont2019,gholaminejad2019anode} can also be seen as an infinite-layer limit of a normalizing flow. Alternatively, Stein variational methods \cite{liu2016stein,liu2017stein,Detommaso2018} provide a nonparametric way of constructing $T$ as a composition of functions lying in a chosen RKHS.

In general, it can be difficult to represent expressive maps in high dimensions. For example, triangular maps on $\mathbb{R}^d$ must describe $d$-variate functions and thus immediately encounter the curse of dimensionality. Similarly, kernel-based methods lose expressiveness in high dimensions \cite{Detommaso2018,Chen2019}. Flow-based methods often increase expressiveness by adding layers, but this is typically performed in an ad hoc or unstructured way, which also requires tuning. 

Here we propose a framework for inference that creates target-informed architectures around \emph{any} class of transport maps or normalizing flows. In particular, our framework uses rigorous \textit{a priori} error bounds to discover and exploit low-dimensional structure in a given target distribution. It also provides a methodology for efficiently solving high-dimensional inference problems via greedily constructed compositions of \emph{structured} low-dimensional maps. 

The impact of our approach rests on two observations. 
First, the coordinate basis in which one expresses a transport map (i.e., $T(x)$ versus $U T(x)$, where $U$ is a rotation on $\R^d$) can 
strongly affect
 the training behavior and final performance of the method. Our framework identifies an ordered basis that best reveals a certain low-dimensional structure in the problem. Expressing the transport map in this basis focuses the expressiveness of the underlying transport class and allows for principled dimension reduction. This basis is identified by minimizing an upper bound on the Kullback--Leibler (KL) divergence between $\pi$ and its approximation, which follows from logarithmic Sobolev inequalities (see \cite{zahm2018certified}) relating the KL divergence to gradients of the target density. 

Second, in the spirit of normalizing flows, we seek to increase the expressiveness of a transport map using repeated compositions. Rather than specifying the length of the flow before training, we increase the length of the flow sequentially. For each layer, we apply the framework above to a \emph{residual} distribution that captures the deviation between the target distribution and its current approximation. We prove weak convergence of this greedy approach to the target distribution under reasonable assumptions. This sequential framework enables efficient layer-wise training of high-dimensional maps, which especially helps control the curse of dimensionality in certain transport classes. As we shall demonstrate empirically, the greedy composition approach can further improve accuracy at the end of training, compared to baseline methods.

%
%




Since Markov chain Monte Carlo (MCMC) methods are also a workhorse of inference, it is useful to contrast them with the variational methods discussed above. In general, these two classes of methods have different computational patterns. In variational inference, one might spend considerable effort to construct the approximate posterior, but afterwards enjoys cheap access to samples and normalized evaluations of the (approximate) target density. How well the approximation matches the true posterior depends on the expressiveness of the approximation class and on one's ability to optimize within this class. 
MCMC, in contrast, requires continual computational effort (even after tuning), but (asymptotically) generates samples from the exact posterior. Yet there is a line of work that uses transport to improve the performance of MCMC methods (\cite{parno2014transport,hoffman2019neutra})---such that even if one desires exact samples, constructing a transport map can be beneficial. We will demonstrate this link in our numerical experiments.


\paragraph{Preliminaries.} We will consider target distributions with densities $\pi$ on $\mathbb{R}^d$ that are differentiable almost everywhere and that can be evaluated up to a normalizing constant. Such a target will often be the posterior of a Bayesian inference problem, e.g., 
$
 \pi(x) \coloneqq p(x\vert y) \propto \mathcal{L}_y(x) \pi_0(x),
$
where $\mathcal{L}_y(x) \coloneqq p(y \vert x)$ is the likelihood function and $\pi_0$ is the prior. We denote the standard Gaussian density on $\mathbb{R}^d$ as $\rho$. We will consider maps $T: \mathbb{R}^d \to \mathbb{R}^d$ that are diffeomorphisms,\footnote{In general $T$ does not need to be a diffeomorphism, but only a
  particular invertible map; see Appendix \ref{app:triangular-maps}
  for more details. The distributions we will consider in this paper, however, fulfill the necessary
  conditions for $T$ to be differentiable almost everywhere.}
and with some abuse of notation, we will write the pushforward density of $\rho$ under $T$ as $T_\sharp\rho(x) \coloneqq \rho \circ T^{-1}(x) \vert \nabla T^{-1}(x) \vert$. We will frequently also use the notion of a \emph{pullback} distribution or density, written as $T^\sharp\pi \coloneqq (T^{-1})_\sharp\pi$.


In \S\ref{sec:LazyMaps} we show how to build a single map in the low-dimensional ``lazy'' format described above, and describe the class of posterior distributions that admit such structure. In \S\ref{sec:LayersLazyMaps} we develop a greedy algorithm for building deep compositions of lazy maps, which effectively decomposes any inference problem into a series of lower-dimensional problems. 
\S\ref{sec:NumericalExamples} presents numerical experiments highlighting the benefits of the lazy framework.
While our numerical experiments employ inverse autoregressive flows \cite{Kingma2016} and polynomial transport maps \cite{Jaini2019,el2012bayesian} as the underlying transport classes, we emphasize that the lazy framework is applicable to any class of transport. 

\section{Lazy maps} \label{sec:LazyMaps}
Given a unitary matrix $U\in\R^{d\times d}$ and an integer $r\leq d$,
let $\mathcal{T}_r(U)$ be the set that contains all the maps
$T:\R^d\rightarrow \R^d$ of the form 
\begin{equation}\label{eq:LazyMap}
 T(z)  =  U \left[\begin{array}{c}
 \tau( z_1,\hdots,z_r) \\
 z_\perp
 \end{array}\right] = 
 U_r \tau( z_1,\hdots,z_r) + U_\perp z_\perp
\end{equation}
for some diffeomorphism $\tau:\R^r\rightarrow\R^r$. Here $U_r\in\R^{d\times r}$ and $U_\perp\in\R^{d\times (d-r)}$ are the matrices containing respectively the $r$ first and the $d-r$ last columns of $U$, and $z_\perp=(z_{r+1},\hdots,z_d)$.
Any map $T\in\mathcal{T}_r(U)$ is called a \emph{lazy map} with rank bounded by $r$, as it is nonlinear only with respect to the first $r$ input variables $z_1,\hdots,z_r$ and the nonlinearity is contained in the low-dimensional subspace $\text{range}(U_r)$. 
The next proposition gives a characterization of all the densities $T_\sharp\rho$ when $T\in\mathcal{T}_r(U)$.

\begin{proposition}[Characterization of lazy maps]\label{prop:CaracterizationOfLazyMaps}
 Let $U\in\R^{d\times d}$ be a unitary matrix and let $r\leq d$. 
 Then for any lazy map $T\in\mathcal{T}_r(U)$, there exists a strictly positive function $f:\R^r\rightarrow\R_{>0}$ such that
 \begin{equation}\label{eq:LazyMap_VS_Ridge}
  T_\sharp\rho (x) = f(U_r^\top x)\rho(x),
 \end{equation}
 for all $x\in\R^d$ where $\rho$ is the density of the standard normal distribution.
 Conversely, any probability density function of the form $f(U_r^\top x)\rho(x)$ admits a representation as in \eqref{eq:LazyMap_VS_Ridge} for some $T\in\mathcal{T}_r(U)$.
\end{proposition}
The proof is given in Appendix \ref{proof:CaracterizationOfLazyMaps}.
By Proposition \ref{prop:CaracterizationOfLazyMaps}, any posterior density $\pi(x)\propto\mathcal{L}_y(x)\pi_0(x)$ with standard Gaussian prior $\pi_0=\rho$ and with likelihood function given by $\mathcal{L}_y(x) \propto f(U_r^\top x)$ can be writen \emph{exactly} as $\pi=T_\sharp\rho$ for some lazy map $T\in\mathcal{T}_r(U)$.
%
%
In particular, posteriors of generalized linear models naturally fall into this class; see Appendix \ref{app:glms} for more details. 
Following \cite[Section 2.1]{zahm2018certified}, the solution $T^\star\in\mathcal{T}_r(U)$  to
$$
  \Dkl( \pi || T^\star_\sharp\rho) = \min_{T \in \mathcal{T}_r(U)} \Dkl( \pi || T_\sharp\rho),
$$
is such that $T^\star_\sharp\rho(x)
=  f^\star(U_r^\top x) \rho(x) $, where $f^\star$ is the
conditional expectation 
$$f^\star(x_r)=\Ex \left [ \frac{\pi(X)}{\rho(X)} \vert U_r^\top X=x_r  \right ] $$
with $X\sim\rho$. 
%

Now that we know the optimal lazy map in $\mathcal{T}_r(U)$, it remains to find a suitable matrix $U$ and rank $r$.
%
In Appendix \ref{proof:KL_Tstar_Decomp} we show that 
\begin{equation}\label{eq:KL_Tstar_Decomp}
 \Dkl( \pi || T^\star_\sharp\rho)
 = \Dkl( \pi || \rho) - \Dkl( (U_r^\top)_\sharp \pi_r || \rho_r ) ,
\end{equation}
where $\rho_r$ is the density of the standard normal distribution on $\R^r$ and $(U_r^\top)_\sharp \pi$ is the density of $U_r^\top X$ with $X\sim\pi$.
Thus, for fixed $r$, minimizing $\Dkl(\pi||T^\star_\sharp\rho)$ over $U$ is the same as finding the most non-Gaussian marginal $(U_r^\top)_\sharp \pi$. Such an optimal $U$ can be difficult to find in practice. The next proposition instead gives a computable \emph{bound} on $\Dkl( \pi || T^\star_\sharp \rho)$, which we will use to construct a $U$ suitable for our algorithm. 
The proof is given in Appendix \ref{proof:LogSobResult}.

\begin{proposition}\label{prop:LogSobResult}
  Let $(\lambda_i,u_i)\in\R_{\geq0}\times\R^d$ be the $i$-th eigenpair of the eigenvalue problem $Hu_i = \lambda_i u_i$ where $H = \int (\nabla\log\frac{\pi}{\rho} )(\nabla\log\frac{\pi}{\rho} )^\top\d\pi$. Let $U=[u_1,\hdots,u_d]\in\R^{d\times d}$ be the matrix containing the eigenvectors of $H$. Then for any $r\leq d$ we have
 \begin{equation}\label{eq:LogSobResult}
  \Dkl( \pi || T^\star_\sharp \rho) \leq \frac{1}{2} (\lambda_{r+1}+\hdots+\lambda_d).
 \end{equation}
\end{proposition}

Proposition \ref{prop:LogSobResult} suggests constructing $U$ as the matrix of eigenvectors of $H$, 
and that a fast decay in the spectrum of $H$ allows a lazy map with low $r$ to accurately represent the true posterior. 
Indeed, one can guarantee $\Dkl( \pi || T^\star_\sharp \rho) < \varepsilon$ by choosing $r$ to be the smallest integer such that the left-hand side of \eqref{eq:LogSobResult} is below $\varepsilon$.
In practice, 
since the complexity of representing and training a transport map may strongly depend on $r$, we can bound $r$ by some $r_{{\text{max}}} \leq d$ associated with a computational budget for constructing $T$.
This procedure is summarized in Algorithm \ref{alg:LazyMap}.


The practical implementation of Algorithm \ref{alg:LazyMap} relies on the computation of $H$. 
Direct Monte Carlo estimation of $H$, however, requires generating samples from $\pi$, which is not feasible in practice.
%
Instead one can use an importance sampling estimate, taking $$H \approx \frac{1}{K}\sum_{k=1}^K \omega_k (\nabla\log\frac{\pi}{\rho}(X_k))(\nabla\log\frac{\pi}{\rho}(X_k))^\top,$$ where $\{X_k\}_{k=1}^K$ are i.i.d.\ samples from $\rho$ and $\omega_k = \frac{\pi(X_k)}{\rho(X_k)}/(\sum_{k'=1}^K \frac{\pi(X_{k'})}{\rho(X_{k'})})$ are self-normalized weights.
This estimate can have significant variance when $\rho$ is a poor approximation to the target $\pi$ (e.g., in the first stage of the greedy algorithm in \S\ref{sec:LayersLazyMaps}). In this case it is preferable to impose $\omega_k=1$, which reduces variance but yields an biased estimator of $H$; instead, it is an unbiased estimator of $H^{\text{B}} = \int (\nabla\log\frac{\pi}{\rho} )(\nabla\log\frac{\pi}{\rho} )^\top\d\rho$. As shown via the error bounds in \cite[Sec.\ 3.3.2]{zahm2018certified} this matrix still provides useful information regarding the construction of $U$. We consider the differences between the two estimators in Appendix~\ref{app:biasofH}.
Also, since the effective sample size (ESS) of the importance sampling estimate can be computed with little extra cost after collecting samples, one can use this ESS to choose whether to use $H$ or $H^{\text{B}}$. 
Other variance reduction methods may also be applicable. For example, simplifications or approximations to the expected outer product of score functions yield natural candidates for control variates.



In constructing a lazy map $T$ of the form \eqref{eq:LazyMap}, one needs to identify a map $\tau:\R^r \to \R^r$ such that $T_\sharp\rho$ approximates the posterior. One can use any transport class to parameterize $\tau$; Appendices
\ref{app:triangular-maps} and \ref{app:iafmaps} detail the particular maps used in our numerical experiments.
%
%
%
In our setting we can only evaluate $\pi$ up to a normalizing constant, and thus it is expedient to minimize the reverse KL divergence
$\Dkl(T_\sharp\rho||\pi) = \Dkl(\rho||T^\sharp\pi)$, as is typical in variational Bayesian methods---which can be achieved by maximizing a Monte Carlo or quadrature approximation of $\Ex_\rho \left [\log T^\sharp \pi \right ]$. This is equivalent to maximizing the evidence lower bound (ELBO) and using the  {reparameterization trick} \cite{kingma2015reparam} to write the expectation over the base distribution $\rho$. 
Details on the numerical implementation of Algorithm \ref{alg:LazyMap} are given in Appendix \ref{app:numerical-algorithms}. We note that the lazy framework works to control the KL divergence in the inclusive direction, while optimizing the ELBO minimizes the KL divergence in the exclusive direction. We show empirically that this computational strategy provides performance improvements in both directions of the KL divergence between the true and approximate posterior, compared to a baseline that does not utilize the lazy framework. 

\begin{algorithm}
	\caption{
		Construction of a lazy map.
	}
	\begin{algorithmic}[1]
		\Procedure{LazyMap}{$\pi$, $\rho$, $\varepsilon$, $r_{{\text{max}}}$}
		\State Compute \label{alg:LazyMap:H}
		$H=\int (\nabla\log\frac{\pi}{\rho} )(\nabla\log\frac{\pi}{\rho} )^\top\d\pi$ 
		\State Solve the eigenvalue problem $Hu_i=\lambda_iu_i$ 
		\State Let $r= r_{{\text{max}}} \wedge \min \{ r\leq d ~: \frac{1}{2}\sum_{i>r}\lambda_{i} \leq \varepsilon \}$ and assemble $U=[u_1,\hdots,u_d]$.    
		\State Find $T$ by solving \label{alg:LazyMap:min} 
		$\min_{T\in\mathcal{T}_r(U)}\Dkl(T_\sharp\rho||\pi)$
		\State\Return lazy map $T$
		\EndProcedure
	\end{algorithmic}\label{alg:LazyMap}
\end{algorithm}

\section{Deeply lazy maps}\label{sec:LayersLazyMaps}

\begin{algorithm}[t]
  \caption{
    Construction of a deeply lazy map
  }
  \begin{algorithmic}[1]
    \Procedure{LayersOfLazyMaps}{$\pi$, $\rho$, $\varepsilon$, $r$, $\ell_{{\text{max}}}$}
    \State Set $\pi_0=\pi$ and $\ell=0$
    \While{$\ell \leq \ell_{{\text{max}}}$ and $\frac{1}{2}\trace(H_{\ell})\geq \varepsilon$}
        \State $\ell\gets\ell+1$
        \State Compute $T_\ell  = $ \textsc{LazyMap}($\pi_{\ell-1}$, $\rho$, $0$, $r$) \Comment{Algorithm \ref{alg:LazyMap}} 
        \State Update $\frak{T}_{\ell}=\frak{T}_{\ell-1}\circ T_\ell$ 
        \State Compute $\pi_\ell = (\frak{T}_{\ell})^\sharp \pi$ 
        \State Compute $H_{\ell}=\int (\nabla\log\frac{\pi_\ell}{\rho} )(\nabla\log\frac{\pi_\ell}{\rho} )^\top\d\pi_\ell$
    \EndWhile
    
    \State\Return $\frak{T}_\ell=T_1\circ\cdots\circ T_\ell$
    \EndProcedure
  \end{algorithmic}\label{alg:LayersOfLazyMaps}
\end{algorithm}

The restriction $r\leq r_{{\text{max}}}$ in Algorithm \ref{alg:LazyMap} helps control the computational cost of constructing the lazy map, but unless a problem admits sufficient lazy structure, $T_\sharp\rho$ may not adequately approximate the posterior. To extend the numerical benefits of the lazy framework to general problems, we consider the ``deeply lazy'' map $\frak{T}_\ell$, a composition of $\ell$ lazy maps:
\[
 \frak{T}_\ell = T_1\circ\hdots\circ T_\ell ,
 \quad T_k \in\mathcal{T}_{r}(U^{k}) ,
\]
where each $T_k$ is a lazy map associated with a different unitary matrix $U^k\in\R^{d\times d}$.
For simplicity we consider the case where each lazy layer $T_k$ has the same rank $r$,
though it is trivial to allow the ranks to vary from layer to layer.
In general, the composition of lazy maps is not itself a lazy map. For example, there exists $U^1\neq U^2$ such that $\frak{T}_2=T_1\circ T_2$ can depend nonlinearly on each input variable and so $\frak{T}_2$ cannot be written as in \eqref{eq:LazyMap}. 



The diagnostic matrix $H$ allows us to build deeply lazy maps in a greedy way.
After $\ell-1$ iterations, the composition of maps
$\frak{T}_{\ell-1}=T_1\circ\hdots\circ T_{\ell-1}$ has been
constructed.
We seek a unitary matrix $U^{\ell}\in\R^{d\times d}$ and a lazy map $T_{\ell}\in\mathcal{T}_r(U^{\ell})$ such that $(\frak{T}_{\ell-1} \circ  T_\ell)_\sharp\rho$ best improves over  $(\frak{T}_{\ell-1})_\sharp\rho$ as an approximation to the posterior. To this end, we define the residual distribution
\[ 
\pi_{\ell-1} = (\frak{T}_{\ell-1})^\sharp \pi,
\]
i.e., the pullback of $\pi$ through the current transport map $\frak{T}_{\ell-1}$.
Note that $\Dkl( \pi||(\frak{T}_{\ell-1} \circ  T_\ell)_\sharp \rho) = \Dkl( \pi_{\ell-1} ||(T_\ell)_\sharp \rho)$. We thus build $T_\ell$ using Algorithm \ref{alg:LazyMap}, replacing the posterior $\pi$ by the residual distribution $\pi_{\ell-1}$.
We then update the transport map to be $\frak{T}_{\ell}=\frak{T}_{\ell-1}\circ T_\ell$ and the residual density $\pi_\ell = (\frak{T}_{\ell})^\sharp \pi$.

We note that applying Proposition \ref{prop:LogSobResult} to $\pi_\ell$ with $r=0$ yields
 \[
 \Dkl(\pi||(\frak{T}_\ell)_\sharp\rho)=\Dkl(\pi_\ell||\rho)\leq \frac{1}{2}(\lambda_1+\cdots+\lambda_d)=\frac{1}{2}\trace(H_{\ell}),
 \]
 where we define the diagnostic matrix at iteration $\ell$ as, 
\[ 
H_{\ell}=\int \left (\nabla\log\frac{\pi_\ell}{\rho} \right ) \left (\nabla\log\frac{\pi_\ell}{\rho}  \right )^\top\d\pi_\ell.
\]
Our framework thus naturally exposes the error bound  $\frac12 \trace{(H_\ell)}$ on the forward KL divergence, which is of {independent} interest and applicable to \emph{any} flow-based method. We refer to this bound as the \emph{trace diagnostic}.

This bound can also be used as a stopping criterion for the greedy algorithm; one can continue adding layers until the bound falls below some desired threshold. 
This construction is summarized in Algorithm \ref{alg:LayersOfLazyMaps}, and details on its numerical implementation are given in Appendix~\ref{app:numerical-algorithms}.


The next proposition gives a sufficient condition on $U^\ell$ to guarantee the convergence of our greedy algorithm. The proof is given in Appendix \ref{proof:GreedyConvergence_condition}.
\begin{proposition}\label{prop:GreedyConvergence_condition}
 Let $U^1,U^2,\hdots$ be a sequence of unitary matrices.
 For any $\ell\geq1$, we let $T_\ell \in\mathcal{T}_r(U^\ell)$ be a lazy map that minimizes $\Dkl( \pi_{\ell-1} ||(T_\ell)_\sharp \rho)$, where $\pi_{\ell-1} = (T_1\circ\hdots\circ T_{\ell-1})^\sharp \pi$.
 If there exists $0< t\leq 1$ such that for any $\ell\geq1$
 \begin{equation}\label{eq:GreedyConvergence_condition}
  \Dkl( (U_r^{\ell \top})_\sharp \pi_{\ell-1} || \rho_r )
  \,\geq\, t  \sup_{\substack{ U\in\R^{d\times d} \\ \text{s.t.} \ UU^\top=I_d}} \Dkl( (U_r^\top)_\sharp \pi_{\ell-1} || \rho_r ),
 \end{equation}
 
 then $(T_1\circ\hdots\circ T_\ell)_\sharp\rho $ converges weakly to $\pi$.
\end{proposition}
Let us comment on the condition \eqref{eq:GreedyConvergence_condition}. Recall that the unitary matrix $U$ that maximizes $\Dkl( (U_r^\top)_\sharp \pi_{\ell-1} || \rho_r )$ is optimal; see \eqref{eq:KL_Tstar_Decomp}.
By \eqref{eq:GreedyConvergence_condition}, the case $t=1$ means that $U^\ell$ is optimal at each iteration. This corresponds to an
\textit{ideal} greedy algorithm. The case $0<t<1$ allows suboptimal
choices for $U^\ell$ without losing the convergence property of the
algorithm. Such a greedy algorithm converges even with a
potentially crude selection of $U^\ell$ that corresponds to a $t$
close to zero. This also is why an approximation to
$H_\ell$ is expected to be sufficient; see Section
\ref{sec:NumericalExamples}. 
We emphasize that condition \eqref{eq:GreedyConvergence_condition} must apply simultaneously to \emph{all} layers for a given $0<t\leq1$. Following \cite{Temlyakov2011}, one could relax this condition
by replacing $t$ with a sequence $(t_\ell)$ that goes to zero sufficiently slowly. This development is left for future work.
%
%
Finally, note that Proposition \ref{prop:GreedyConvergence_condition} does not require any constraints on $r$, so we have convergence even with $r=1$, where each layer only acts on a single direction at a time. 


\begin{figure}[t]
  \centering
  \begin{minipage}{.47\linewidth}
    \begin{subfigure}[b]{\textwidth}
      \includegraphics[width=0.9\textwidth]{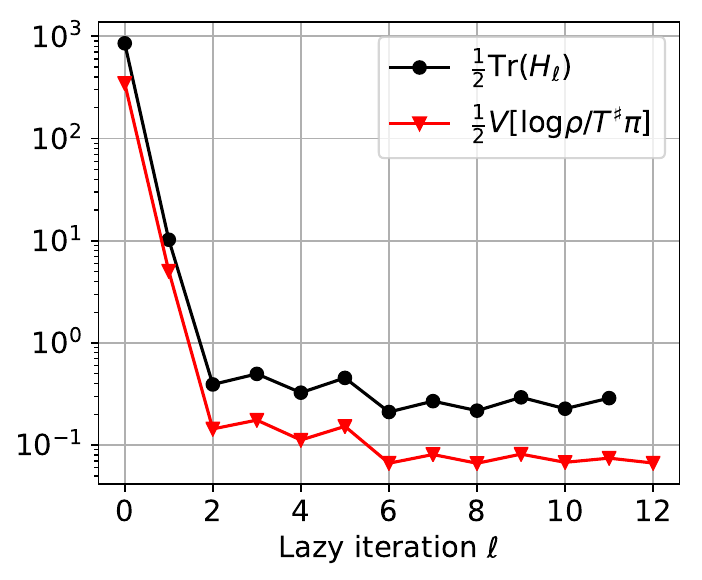}
      \caption{Convergence rate}
      \label{fig:ex:banana:convergence}
    \end{subfigure}
  \end{minipage}
  \hspace{3pt}
  \begin{minipage}{.5\linewidth}
    \begin{subfigure}[b]{0.28\textwidth}
      \includegraphics[width=\textwidth]{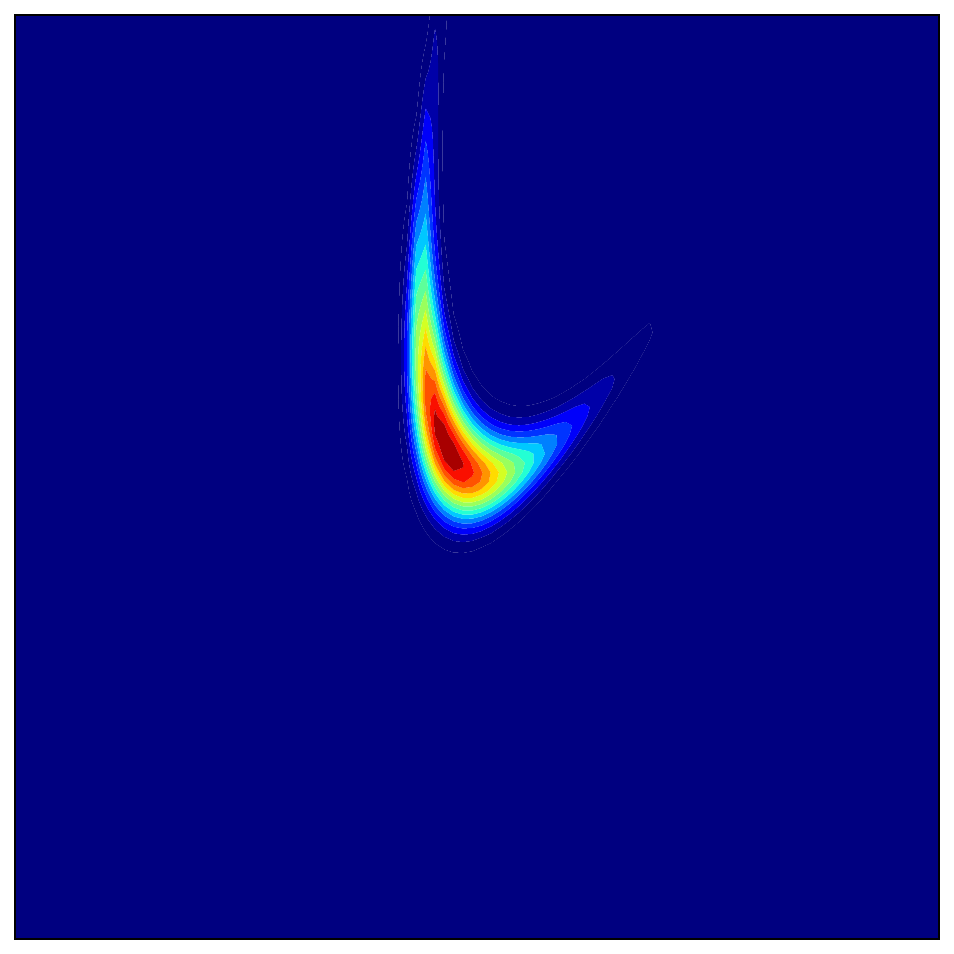}
      \caption{Target $\pi$}
      \label{fig:ex:banana:target}
    \end{subfigure}
    \hspace{3pt}
    \begin{subfigure}[b]{0.28\textwidth}
      \includegraphics[width=\textwidth]{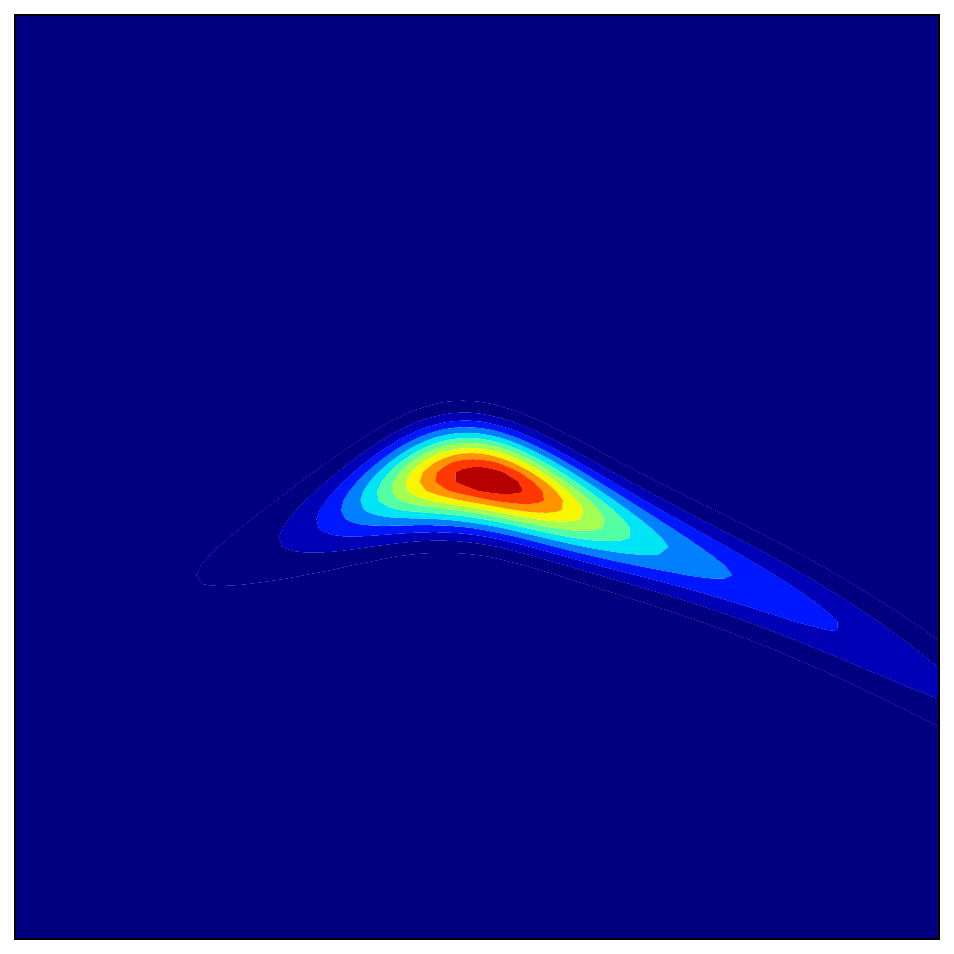}
      \caption{$\frak{T}_1^\sharp\pi$}
      \label{fig:ex:banana:pullback-1}
    \end{subfigure}
    \hspace{3pt}
    \begin{subfigure}[b]{0.28\textwidth}
      \includegraphics[width=\textwidth]{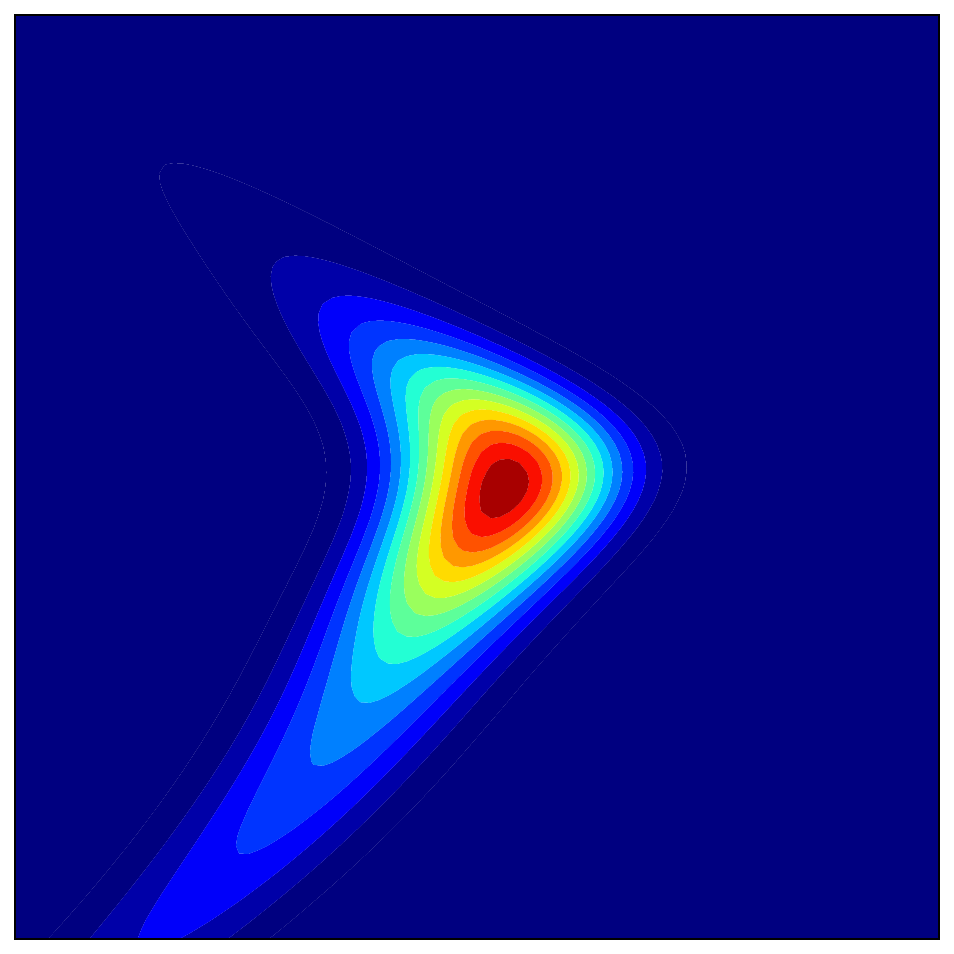}
      \caption{$\frak{T}_2^\sharp\pi$}
    \end{subfigure}
    \\[5pt]
    \begin{subfigure}[b]{0.28\textwidth}
      \includegraphics[width=\textwidth]{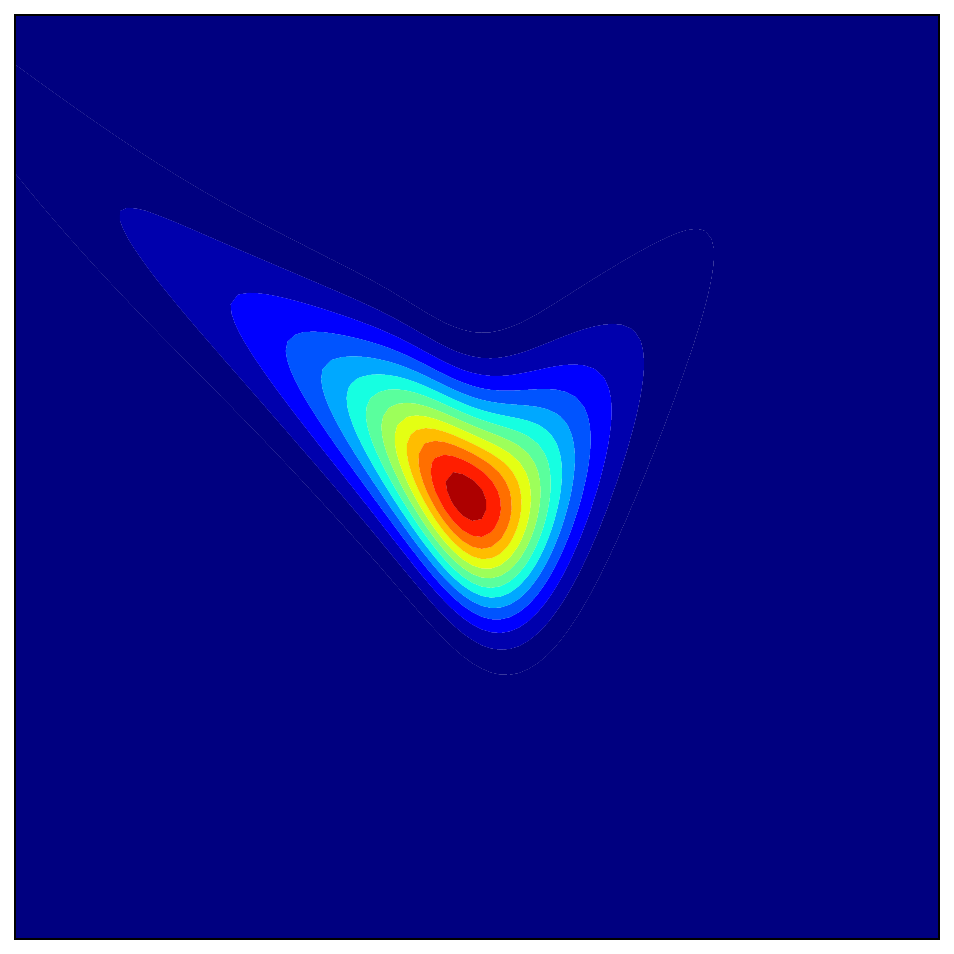}
      \caption{$\frak{T}_3^\sharp\pi$}
    \end{subfigure}
    \hspace{3pt}
    \begin{subfigure}[b]{0.28\textwidth}
      \includegraphics[width=\textwidth]{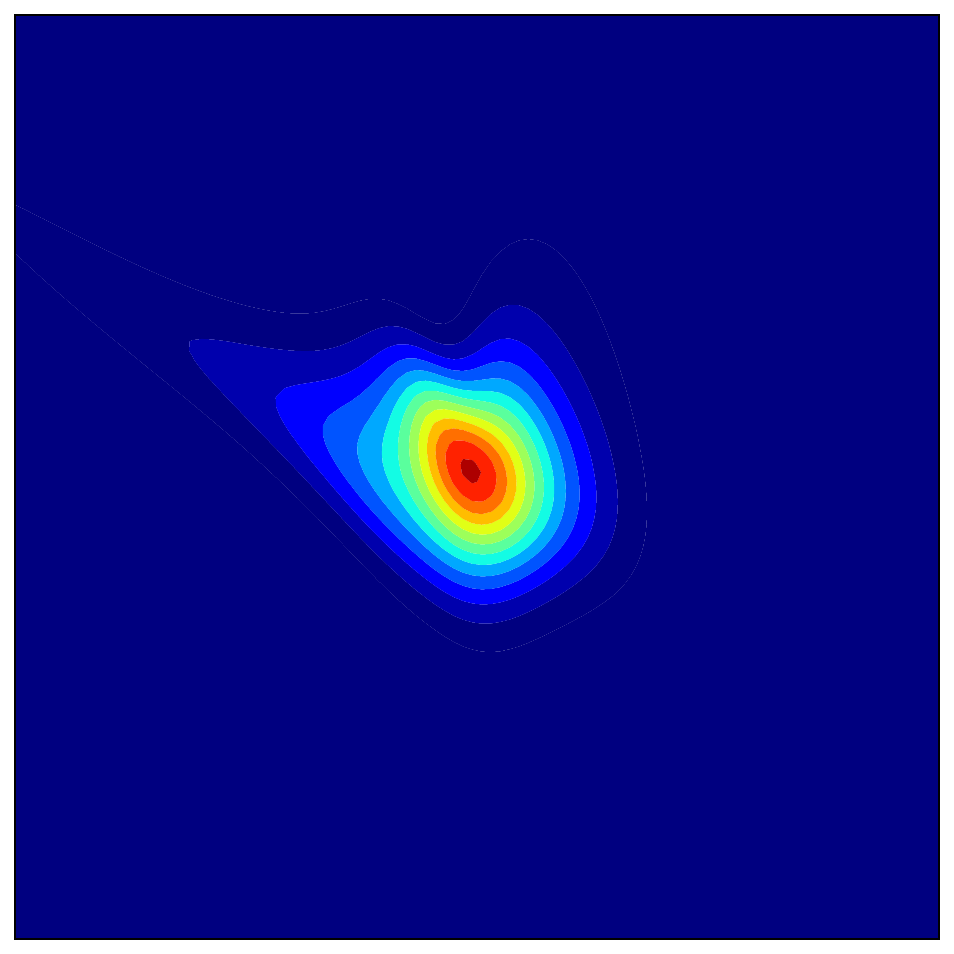}
      \caption{$\frak{T}_5^\sharp\pi$}
    \end{subfigure}
    \hspace{3pt}
    \begin{subfigure}[b]{0.28\textwidth}
      \includegraphics[width=\textwidth]{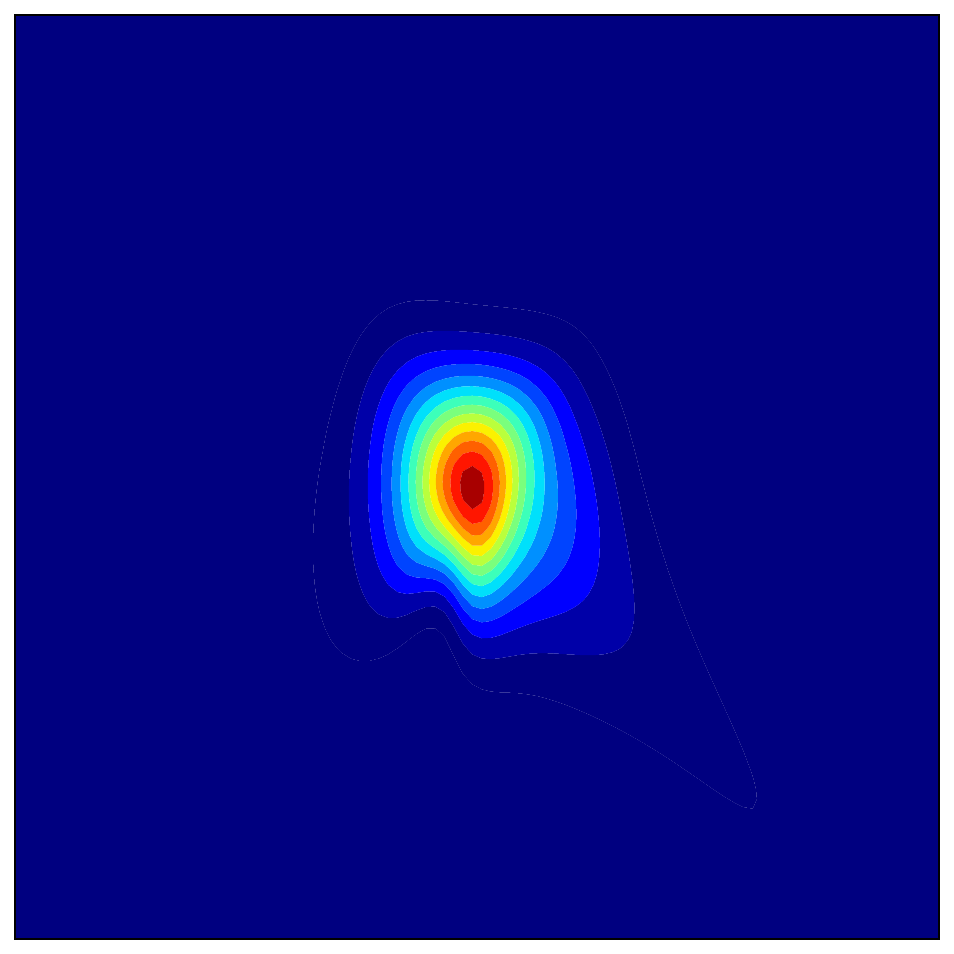}
      \caption{$\frak{T}_8^\sharp\pi$}
      \label{fig:ex:banana:pullback-8}
    \end{subfigure}
  \end{minipage}
  \caption{
    Convergence of the algorithm for the approximation of the rotated
    banana distribution. (\subref{fig:ex:banana:convergence}) Decay of the bound
    $\frac{1}{2}\trace(H^{\text{B}}_\ell)$ on the KL-divergence
    $\Dkl(\pi \Vert (\frak{T}_{\ell})_\sharp \rho)$ and the variance diagnostic $\frac{1}{2}\mathbb{V}_\rho[\log\rho / \frak{T}_\ell^\sharp \pi]$.
    (\subref{fig:ex:banana:target}) The target density $\pi$.
    (\subref{fig:ex:banana:pullback-1}--\subref{fig:ex:banana:pullback-8})
    The target distribution is progressively Gaussianized by the maps
    $\frak{T}_\ell$.
  }
  \label{fig:ex:banana}
\end{figure}

\section{Numerical examples}\label{sec:NumericalExamples}
\label{sec:numerical-examples}
We present numerical demonstrations of the lazy framework as follows. We first illustrate Algorithm \ref{alg:LayersOfLazyMaps} on a 2-dimensional toy example, where we show the progressive Gaussianization of the posterior using a sequence of 1-dimensional lazy maps.
We then demonstrate the benefits of the lazy framework (Algorithms \ref{alg:LazyMap} and \ref{alg:LayersOfLazyMaps}) in several challenging inference problems. We consider Bayesian logistic regression and a Bayesian neural network, and compare the performance of a baseline transport map to lazy maps using the same underlying transport class. We measure performance improvements in four ways: (1) the final ELBO achieved by the transport maps after training; (2 and 3): the final trace diagnostics $\frac{1}{2}\trace(H^{\text{B}}_{\ell})$ and $\frac{1}{2}\trace(H_{\ell})$, which bound the error $\Dkl(\pi|| (\frak{T}_{\ell})_\sharp \rho)$; and (4) the \textit{variance diagnostic}
$\frac{1}{2}\mathbb{V}_\rho[\log\rho / \frak{T}_\ell^\sharp \pi]$, which is an asymptotic approximation of $\Dkl( (\frak{T}_{\ell})_\sharp \rho ||\pi)$ as $(\frak{T}_{\ell})_\sharp \rho \rightarrow \pi$ (see \cite{el2012bayesian}).
Finally, we highlight the advantages of greedily training lazy maps in a nonlinear problem defined by a high-dimensional elliptic partial differential equation (PDE), often used for testing high-dimensional inference methods \cite{clm_2014, beskos2017geometric, stuart2010inverse}. Here, the lazy framework is needed to make variational inference tractable by controlling the total number of map parameters. We also illustrate the utility of such flows in preconditioning Markov chain Monte Carlo (MCMC) samplers \cite{parno2014transport,hoffman2019neutra}, or equivalently as a way of de-biasing the variational approximation on these three problems.

Numerical examples are implemented
\footnote{Code for the numerical examples can be found at \url{https://github.com/MichaelCBrennan/lazymaps} and  \url{http://bit.ly/2QlelXF}. Data for \S \ref{sec:examples:elliptic}, \ref{app:examples:log-cox}, and \ref{app:examples:timoshenko} can be downloaded at \url{http://bit.ly/2X09Ns8}, \url{http://bit.ly/2HytQc0} and \url{http://bit.ly/2Eug5ZR.} }
both in the \texttt{TransportMaps} framework \cite{transportmaps} and using the TensorFlow probability library \cite{dillon2017tensorflow}. The PDE considered in \ref{sec:examples:elliptic} is discretized and solved using the \texttt{FEniCS} \cite{LoggWells2010a} and \texttt{dolfin-adjoint} \cite{Farrell2013} packages.
%


\subsection{Illustrative toy example}\label{sec:examples:banana}

We first apply the algorithm on the standard problem of approximating
the rotated banana distribution $Q_\sharp\pi_{X_1,X_2}$ defined by
$X_1\sim\mathcal{N}(0.5,0.8)$ and $X_2\vert X_1 \sim\mathcal{N}(X_1^2,0.2)$,
and where $Q$ is a random rotation.
We restrict ourselves to using a composition of
rank-1 lazy maps. We consider degree 3 polynomial maps as the underlying transport class.
We use Gauss quadrature rules of order 10 for the discretization of the
KL divergence and the approximation of $H^{\text{B}}_\ell$
($m=121$ in Algorithm \ref{alg:compute-H-rho} and \ref{alg:compute-map}).
Figure \ref{fig:ex:banana:target} shows the target distribution $\pi\coloneqq\pi_{X_1,X_2}$.
Figure \ref{fig:ex:banana:convergence} shows the convergence of the algorithm
both in terms of the trace diagnostic $\frac{1}{2}\trace(H^{\text{B}}_\ell)$ and in terms of the
variance diagnostic. After two iterations the algorithm
has explored all directions of $\mathbb{R}^2$, leading to a fast improvement.
The convergence stagnates once the trade-off between the complexity of the underlying transport class and the accuracy of the quadrature
has been saturated. Figures \ref{fig:ex:banana:pullback-1}--\subref{fig:ex:banana:pullback-8}
show the progressive Gaussianization of the residual distributions $\frak{T}_\ell^\sharp\pi$ for different iterations $\ell$.



\subsection{Bayesian logistic regression}
\label{sec:examples:blr}

We now consider a high-dimensional Bayesian logistic regression problem using the UCI Parkinson's disease classification data \cite{parkinsonsUCI}, studied in \cite{sakar2019comparative}. We consider the first $500$ provided attributes consisting mainly of patient audio extensions. This results in a $d=500$ dimensional inference problem. We choose a relatively uninformative prior of $\sfN(0,10^2 I_d)$.

Here we consider inverse autoregressive flows (IAFs) \cite{Kingma2016} for the underlying transport class. Details on the IAF structure, our choice of hyper-parameters, and training procedure 
are in Appendix \ref{app:iafmaps}.

As noted in \S\ref{sec:LazyMaps} and shown in Appendix \ref{app:glms}, generalized linear models can admit an exactly lazy structure, where the lazy rank $r$ of the posterior is bounded by the number of observations. We demonstrate this by first considering a small subset of $20$ observations. Given a sufficiently expressive underlying transport class, a single lazy map of rank $r = 20$ can exactly capture the posterior.
We compare three transport maps: a baseline IAF map; $U$-IAF, which is a 1-layer lazy map with rank $r = d = 500$ expressed in the computed basis $U$; and $U_{r}$-IAF, which is a 1-layer lazy map of rank $r = 20$. The diagnostic matrix $H^{\text{B}}$ yielding this basis was computed using $500$ standard normal samples. Results are summarized in Table \ref{ex:tab:bnnblr}. We see improved performance in each of the lazy maps compared to the baseline. We also note that $U_r$-IAF outperforms $U$-IAF in each metric. While the number of flow parameters in $U$-IAF is greater than in $U_r$-IAF, the latter only acts on a $20$ dimensional subspace, and in fact has a higher ratio of map parameters to active dimensions. This highlights a key benefit of the lazy framework: the ability to focus the expressiveness of a transport map along particular subspaces important to the capturing the posterior.
\begin{wrapfigure}{R}{0.43\columnwidth}
	\centering
	\vspace{-3ex}
	\includegraphics[width=6cm]{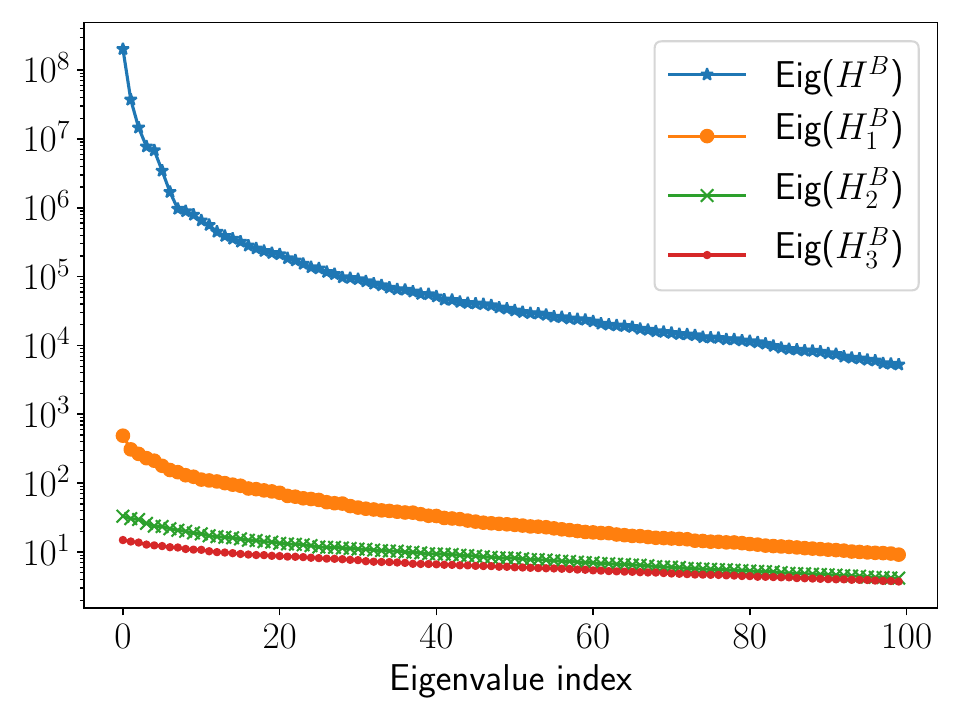}
	\caption{Leading eigenvalues of the diagnostic matrices $H^{\text{B}}_\ell$ for the G3-IAF map applied to the full rank logistic regression problem. The spectrum flattens and falls as the approximation to the posterior improves. \vspace{-1em}}
	\label{fig:blr_fullrank_eigenvalues}
\end{wrapfigure}
	
Next we consider a full rank Bayesian logistic regression problem using $605$ observations. Here we compare a baseline IAF; $U$-IAF defined as before; and a $3$-layer lazy map trained via the greedy  Algorithm \ref{alg:LayersOfLazyMaps}, denoted G3-IAF. In G3-IAF, each layer has rank $r = 200$. Results are summarized in Table \ref{ex:tab:bnnblr}, and again we see improvements in each of the performance metrics compared to the baseline IAF. Recall that the basis $U$ relates to a bound on the inclusive KL direction, while the objective function for map training within a layer optimizes the exclusive KL direction. Empirically we see benefits in metrics relating to both directions. Interestingly, we observe that $U$-IAF achieves the greatest ELBO while G3-IAF achieves the lowest trace diagnostics. This suggests that using a larger number of lazy layers tends to lead to improvements to the inclusive KL divergence. Also, though we chose to use the same number of training iterations in each case, we observe that training of the lazy maps converges more quickly; see Appendix \ref{app:example:blr} for more details.

As discussed in the introduction, a powerful use case for transport maps is the ability to precondition an MCMC method as described in \cite{parno2014transport,hoffman2019neutra,peherstorfer2019transport}, i.e., using the computed map to improve the posterior geometry. Applying Hamiltonian Monte Carlo \cite{neal2011mcmc} to the full rank Bayesian logistic regression problem (in particular, sampling the pullback $\frak{T}_\ell^\sharp\pi$ where $\frak{T}_\ell$ is the learned $U$-IAF map), we achieve worst, best, and average component-wise effective sample sizes of $0.42\%$, $1.2\%$, and $0.87\%$, compared to $0.056\%$, $0.12\%$, and $0.065\%$ without a transport map (sampling the target $\pi$ directly). Note that applying $\frak{T}_\ell$ to MCMC samples from the pullback yields asymptotically exact samples from $\pi$.  Three leapfrog steps were used in the HMC proposal, and the step sizes were chosen adaptively during the burn-in period of the chains to obtain acceptance rates between $70\%$ and $90\%$ \cite{andrieu2008tutorial,beskos2013optimal,betancourt2014optimizing}.

%
%
\begin{table}
	\caption{\label{ex:tab:bnnblr}Result summaries for the Bayesian logistic regression and Bayesian neural network examples. Values reported are the median and (interquartile range) across 10 trials with randomized initialization. Best performance is bolded. $\Delta$ ELBO computed using the median of the baseline.\vspace{.5em}}

\begin{tabular}{ c*{4}{c} }
	Map &  $\Delta$ ELBO* $(\uparrow)$  & Variance diagnostic $(\downarrow)$ & $\trace(H_\ell^{\text{B}})/2$ $(\downarrow)$ & $\trace(H_\ell)/2$ $(\downarrow)$ \\ \hline
	\multicolumn{5}{c}{Low rank Bayesian logistic regression} \\
	\hline
	Baseline IAF    & --  & $23.8$ ($2.59$)& $121$ ($17.2$)  & $56$ ($17.6$) \\
	$U$-IAF    & $6.62$ ($0.368$)  & $7.68$ ($2.07$)& $38.1$ ($7.67$)  & $20.5$ ($10.3$) \\
	$U_r$-IAF  & $\boldsymbol{11.1}$ ($0.172$)  & $\boldsymbol{1.58}$ ($0.433$)& $\boldsymbol{9.85}$ ($2.78$)  & $\boldsymbol{5.95}$ ($1.53$) \\
	\hline
	\multicolumn{5}{c}{Full rank Bayesian logistic regression} \\
	\hline	
	Baseline IAF &--  & $223$ ($27.9$) & $980$ ($84.3$)  & $338$ ($177$)\\
	$U$-IAF    & $\boldsymbol{27.6}$ ($0.529$)  & $141$ ($5.21$              & $651$ ($21.5$)  & $252$ ($132$) \\
	G3-IAF    & $1.23$ ($1.38$)                 & $\boldsymbol{122}$ ($14.1$)& $\boldsymbol{526}$ ($30.6$)  & $205$ ($118$) \\
	\hline
	\multicolumn{5}{c}{Bayesian neural network} \\
	\hline
	Baseline Affine & --  & $1.6\text{e}4$ ($5.8\text{e}4$)& $3.5\text{e}5$ ($6.9\text{e}5$)  & $960$ ($1.0\text{e}3$) \\
	G3-Affine   & $\boldsymbol{47.7}$ ($2.33$)  & $\boldsymbol{97.5}$ ($6.47$)& $\boldsymbol{1.06\text{e}3}$ ($56.2$)  & $\boldsymbol{606}$ ($201$)  \\ \hline 
\end{tabular}
	
\end{table}

	
 
\subsection{Bayesian neural network}
\label{sec:examples:bnn}

We now consider a Bayesian neural network, also in \cite{liu2016stein,Detommaso2018}, trained on the UCI yacht hydrodynamics data set \cite{yachtrockUCI}. Our inference problem is 581-dimensional, given a network input dimension of 6, one hidden layer of dimension 20, and an output layer of dimension 1. We use sigmoid activations in the input and hidden layer, and a linear output layer. Model parameters are endowed with independent Gaussian priors with zero mean and variance 100. Further details are in Appendix~\ref{app:example:bnn}.

Here we consider affine maps as the underlying class of transport. This yields Gaussian approximations to the posterior distribution in both the lazy and baseline cases. We compare a baseline affine map and G3-affine, denoting a 3-layer lazy map where each layer has rank $r = 200$. The diagnostic matrices $H^{\text{B}}_\ell$ are computed using $581$ standard normal samples. We note improvements in each of the performance metrics using the lazy framework, summarized in Table~\ref{ex:tab:bnnblr}. We also note a 64\% decrease in the number of trained flow parameters in G3-affine, relative to the baseline case 
(from $338142$ to $120600$). 

Similarly to \S\ref{sec:examples:blr}, we compare the performance of HMC applied with and without transport map preconditioning. We achieve worst, best, and average component-wise ESS of $0.073 \%$, $1.2\%$, and $0.56\%$ using the learned $G3$-Affine map, compared to $0.047\%$, $0.14\%$, and $0.06\%$ without a transport map. Here five leapfrog steps were used in the HMC proposal, and the step sizes in each case were picked adaptively as before.


\subsection{High-dimensional elliptic PDE inverse problem}
\label{sec:examples:elliptic}

\begin{figure}
  \centering
  \begin{subfigure}[b]{.31\textwidth}
    \includegraphics[width=\textwidth]{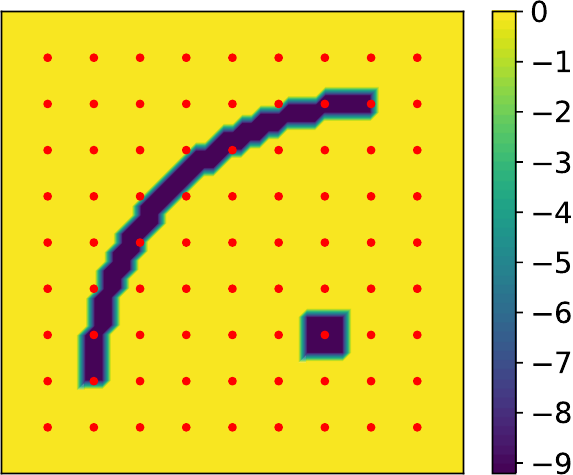}
    \caption{Data generating field $\kappa$}
    \label{fig:ex:elliptic:true-kappa}
  \end{subfigure}
  \hspace{2pt}
  \begin{subfigure}[b]{.31\textwidth}
    \includegraphics[width=\textwidth]{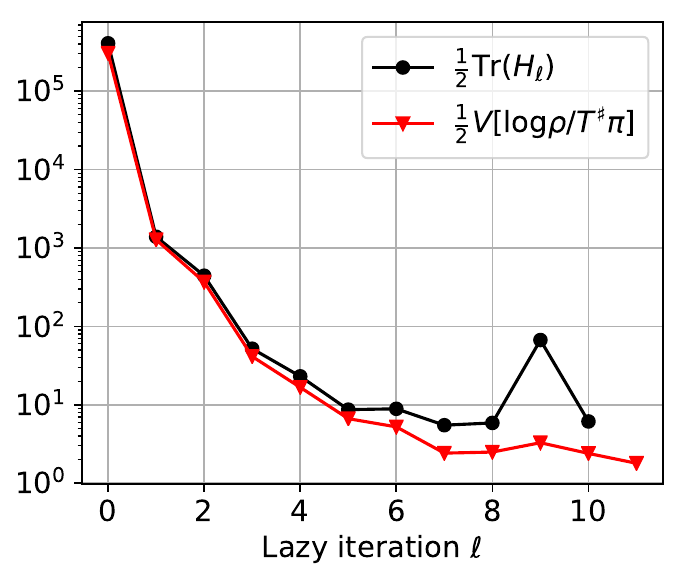}
    \caption{Convergence}
    \label{fig:ex:elliptic:convergence}
  \end{subfigure}
  \hspace{2pt}
  \begin{subfigure}[b]{.31\textwidth}
    \centering
    \includegraphics[width=\textwidth]{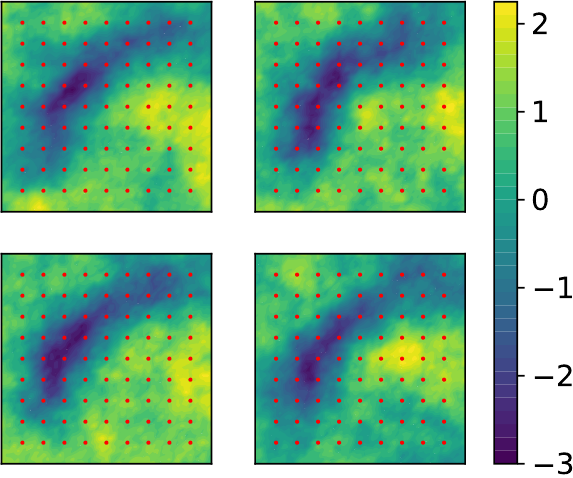}
    \caption{Realizations of $\kappa\sim \pi(\kappa\vert\yb^\star)$}
    \label{fig:ex:elliptic:log-kappa-realizations}
  \end{subfigure}
  \caption{
    Application of Algorithm~\ref{alg:LayersOfLazyMaps} to an elliptic PDE with
    unknown diffusion coefficient.
    (\subref{fig:ex:elliptic:true-kappa}) The data-generating field $\kappa$. 
    (\subref{fig:ex:elliptic:convergence}) Convergence of the trace error bound and variance diagnostic with greedy iterations.
    (\subref{fig:ex:elliptic:log-kappa-realizations})
    Draws from the 2601-dimensional posterior distribution. 
  }
  \label{fig:ex:elliptic}
\end{figure}

We consider the problem of estimating the 
diffusion coefficient $e^{\kappa(\xb)}$
of an elliptic PDE from sparse
observations of the field $u(\xb)$ solving
\begin{equation}
  \label{eq:elliptic}
  \begin{cases}
    \nabla \cdot (e^{\kappa(\xb)} \nabla u(\xb)) = 0, \ \  \text{for } \xb \in \Dc\coloneqq[0,1]^2 \;,\\
    u(\xb) = 0\  \text{for } \xb_1 = 0,\  u(\xb) = 1 \ \text{for } \xb_1 = 1, \ 
    \frac{\partial u(\xb)}{\partial {\bm n}} = 0\  \text{for } \xb_2 \in \{0, 1\} \;.
  \end{cases} 
\end{equation}
This PDE is discretized using finite elements
over a uniform mesh of $51\times 51$ nodes, leading to $d=2601$ degrees of freedom.
We denote by $\kappab$ the discretized version of the log-diffusion
coefficient over this mesh.
Let $\Fc$ be the map from the parameter
$\kappab$ to $n=81$ values of $u$
collected at the locations shown in Figure \ref{fig:ex:elliptic:true-kappa}.
Observations follow the model $\yb = \Fc(\kappab) + \epsilon$,
where $\epsilon\sim\Nc(0,\Sigma_{\text{obs}})$ and
$\Sigma_{\text{obs}}\coloneqq 10^{-3}I_d$. 
The coefficient $\kappab$ is endowed with a Gaussian prior
$\Nc(0,\Sigma)$ where $\Sigma$ is the covariance of an
Ornstein–Uhlenbeck process.
For the observations $\yb^\star$ associated to the parameter $\kappab^\star$
shown in Figure \ref{fig:ex:elliptic:true-kappa},
our target distribution is
$\pi(\zb) \propto \Lc_{\yb^\star}(\zb) \rho(\zb)$,
where $\kappab = \Sigma^{1/2}\zb$.


We greedily train a deeply lazy map using Algorithm~\ref{alg:LayersOfLazyMaps}, using triangular polynomial maps as the underlying transport (see Appendix~\ref{app:triangular-maps}).
Expectations appearing in the
algorithm are discretized with $m=500$ Monte Carlo samples.
To not waste
work in the early
iterations, we use affine maps of rank $r=4$ for iterations
$\ell=1,\ldots,5$.
Then we switch to polynomial maps of degree $2$ and rank $r=2$
for the remaining iterations. This reflects the flexibility of the lazy framework; changes to the underlying transport class and the lazy rank of each layer are simple to implement. The algorithm is terminated when it stagnates after exhausting the expressiveness of the underlying transport class, and the precision of approximating the objective using $m$ samples; see Figure \ref{fig:ex:elliptic:convergence}.
Randomly drawn realizations of
$\kappab$ in Figure \ref{fig:ex:elliptic:log-kappa-realizations}
resemble the generating field.

This elliptic PDE is a challenging benchmark problem for high-dimensional inference \cite{clm_2014, beskos2017geometric, stuart2010inverse}. We note that the final map is a \emph{sparse} degree-$32$ polynomial that acts nonlinearly on all  $2061$ degrees of freedom. Without imposing structure, the curse of dimensionality would render the solution of this problem using polynomial transport maps completely intractable \cite{trefethen2017cubature}. For instance, a na\"{i}ve total-degree parameterization of just the final component of the map would contain ${2061 + 32 \choose 32} \approx 5.5 \times 10^{70}$ parameters. 
We can confirm the quality of the posterior approximation and demonstrate a further application of transport by using MCMC to sample the pullback $\frak{T}_\ell^\sharp\pi$. 
We do so using %
\emph{preconditioned Crank-Nicolson} (\texttt{pCN}) MCMC \cite{Cotter2013} (a state-of-the-art algorithm for PDE problems, with dimension-independent convergence rate) with a step size parameter $\beta=0.5$. The acceptance rate is $28.2\%$ with the worst, best, and average effective sample sizes \cite{Wolff2004} being $0.2\%$, $2.6\%$, and $1.5\%$ of the complete chain. For comparison, a direct application of \texttt{pCN} with the same $\beta$ leads
to an acceptance rate under $0.4\%$ and an effective sample size that cannot be reliably computed.
More details are in Appendix \ref{app:examples:elliptic}.

\section{Conclusions}
We have presented a framework for creating target-informed architectures for transport-based variational inference.
Our approach uses a rigorous error bound to identify low-dimensional structure in the target distribution and focus the expressiveness of the transport map or flow on an important subspace.
%
%
We also introduce and analyze a greedy algorithm for building deep compositions of low-dimensional maps that can iteratively approximate general high-dimensional target distributions. Empirically, these methods improve the accuracy of inference, accelerate training, and control the complexity of flows to improve tractability. Ongoing work will consider constructive tests for further varieties of underlying structure in inference problems, and their implications on the structure of flows.






\cleardoublepage  
\section*{Broader Impact}

\paragraph{Who may benefit from this research?}
We believe users and developers of approximate inference methods will benefit from our work.
Our framework works as an ``outer wrapper'' that can improve the effectiveness of any flow-based variational inference method by guiding its structure. We hope to make expressive flow-based variational inference more tractable, efficient, and broadly applicable, particularly in high dimensions, by developing automated tests for low-dimensional structure and flexible ways to exploit it.
The trace diagnostic developed in our work rigorously assesses the quality of transport/flow-based inference, and may be of independent interest.

\paragraph{Who may be put at disadvantage from this research?}
We don't believe anyone is put at disadvantage due to this research.

\paragraph{What are the consequences of failure of the system?}
We specifically point out that one contribution of this work is identifying when a poor posterior approximation has occurred. A potential failure mode of our framework would be inaccurate estimation of the diagnostic matrix $H$ or its spectrum, suggesting that the approximate posterior is more accurate than it truly is. However, computing the eigenvalues or trace of a symmetric matrix, even one estimated from samples, is a well studied problem. And numerical software guards against poor eigenvalue estimation or at least warns if this occurs. We believe the theoretical underpinnings of this work make it robust to undetected failure. 

\paragraph{Does the task/method leverage biases in the data?}
We don't believe our method leverages data bias. As a method for variational inference, our goal is to accurately approximate a posterior distribution. It is very possible to encode biases for/against a particular result in a Bayesian inference problem, but that occurs at the level of modeling (choosing the prior, defining the likelihood) and collecting data, not at the level of approximating the posterior.

\begin{ack}

This work was supported in part by the US Department of Energy, Office
of Advanced Scientific Computing Research, AEOLUS (Advances in
Experimental Design, Optimal Control, and Learning for Uncertain
Complex Systems) project. The authors also gratefully acknowledge
support from the Inria associate team UNQUESTIONABLE.
\end{ack}

\bibliographystyle{abbrv}
\bibliography{thesisBib}

\cleardoublepage  
\appendix 
\section{Proofs}

\subsection{Proof of Proposition \ref{prop:CaracterizationOfLazyMaps}}\label{proof:CaracterizationOfLazyMaps}

We first show that for any $T\in\mathcal{T}_r(U)$, there exists a $f:\R^r\rightarrow\R_{>0}$ such that \eqref{eq:LazyMap_VS_Ridge} holds. Let $T\in\mathcal{T}_r(U)$. Because $T$ is a diffeomorphism we have $T_\sharp\rho (x) = \rho(T^{-1}(x)) \text{det}(\nabla T^{-1}(x))$. The inverse of $T$ is given by
 \[
 T^{-1}(x) = \begin{pmatrix} \tau^{-1}(U_r^\top x) \\ U_\perp^\top  x \end{pmatrix},
 \]
and so 
\[\text{det}(\nabla T^{-1}(x))=\text{det}(\nabla \tau^{-1}(U_r^\top x)).\]
Recalling $\rho(x)\propto\exp(-\frac{1}{2}\|x\|_2^2)$, we have that 
\[
\rho(T^{-1}(x))\propto \rho(x)\exp\left(-\frac{1}{2}\|\tau^{-1}(U_r^\top x)\|_2^2 + \frac{1}{2}\|U_r^\top x\|_2^2\right),
\]
which yields the result of \eqref{eq:LazyMap_VS_Ridge} by defining 
\[
f(U_r^\top x)=\exp\left(-\frac{1}{2}\|\tau^{-1}(U_r^\top x)\|_2^2 + \frac{1}{2}\|U_r^\top x\|_2^2\right)\text{det}(\nabla \tau^{-1}(U_r^\top x)).
\]
 
Now we show that for any function $f:\R^r\rightarrow\R_{>0}$ there exists a lazy map $T\in\mathcal{T}_r(U)$ such that \eqref{eq:LazyMap_VS_Ridge} holds. Let $f:\R^r\rightarrow\R_{>0}$. Denote by $\rho_r$ (resp.\ $\rho_\perp$) the density of the standard normal distribution on $\R^r$ (resp. $\R^{d-r}$). Let $\tau:\R^r\rightarrow\R^r$ be a map that pushes forward $\rho_r$ to $\pi_r$, where $\pi_r$ is the probability density on $\R^r$ defined by $\pi_r(y_r) \propto f(y_r)\rho_r(y_r)$. Such a map always exists because the support of $\pi_r$ (and of $\rho_r$) is $\R^r$ (see \cite{villani2008optimal} for details).
Consider the map $Q:\R^d\rightarrow\R^d$ defined by 
\[
  Q(z)=\begin{pmatrix} \tau(z_1,\hdots,z_r)\\z_\perp\end{pmatrix}.
 \]
 Because $\rho=\rho_r \otimes \rho_\perp$, we have $Q_\sharp\rho(y)=\tau_\sharp\rho_r (y_r) \rho(y_\perp) \propto f(y_r)\rho(y)$. Finally, the lazy map 
 \[T(z) = U_r\tau(z_1,\hdots,z_r)+U_\perp z_\perp=UQ(z)\]
  satisfies 
\[ T_\sharp\rho(z) = U_\sharp( Q_\sharp \rho )(z) \propto f( (U^\top z)_r )\rho(U^\top z) \propto f( U_r^\top z )\rho(z).\]
 
  This concludes the proof.

\subsection{Proof of Relation \eqref{eq:KL_Tstar_Decomp}} \label{proof:KL_Tstar_Decomp}

We can write
\begin{align*}
 \Dkl( \pi || T^\star_\sharp\rho)
 &= \Ex_\pi[\log(\pi/T^\star_\sharp\rho)] = \Ex_\pi[\log(\pi/\rho)] - \Ex_\pi[\log(T_\sharp^\star\rho/\rho)]\\
 &= \Dkl( \pi || \rho) - \int \log\left(f^\star(U_r^\top x)\right)\pi(x) \d x \\
 &= \Dkl( \pi || \rho) - \int \log\left(f^\star(x_r)\right)\pi_r(x_r) \d x_r,
\end{align*}
where $\pi_r(x_r)=(U_r^\top)_\sharp\pi(x_r)$ is the marginal posterior.
To complete the result, we must show that $f^\star(x_r) = \pi_r(x_r)/\rho_r(x_r)$.
By definition of $f^\star$ we have
\begin{align*}
 f^\star(x_r)
 = \int \frac{\pi(x_r+x_\perp)}{\rho(x_r+x_\perp)}   \rho(x_\perp) \d x_\perp
 = \int \frac{\pi(x_r+x_\perp)}{\rho_r(x_r)\rho_\perp(x_\perp)}   \rho(x_\perp) \d x_\perp
 = \frac{\int \pi(x_r+x_\perp)\d x_\perp}{\rho_r(x_r)}
 = \frac{\pi_r(x_r)}{\rho_r(x_r)},
\end{align*}
which concludes the proof.

\subsection{Proof of Proposition \ref{prop:LogSobResult}}\label{proof:LogSobResult}

 Corollary 1 in \cite{zahm2018certified} allows us to write 
 \[
\Dkl( \pi(x) || f^*(U_r^\top  x)\rho(x)) \leq \frac{1}{2}\trace\left[(I_d-U_rU_r^\top )H(I_d-U_rU_r^\top ) \right].
\]
This result follows from a more general \emph{subspace logarithmic Sobolev inequality}, a result that applies to any given projector $P_r \in \R^{d \times d}$ and bounds expectations of the form 
$$
\mathbb{E}_\pi\left[h^2\log\left(\frac{h^2}{\mathbb{E}_\pi[h^2|\sigma(P_r)]} \right)\right],$$ where $\sigma(P_r)$ denotes the $\sigma$-algebra generated by $P_r$ and $h$ is a continuously differentiable function. Here we take $P_r = U_r U_r^\top $, the projector onto the subspace spanned by the first $r$ eigenvectors of $H$. The function $h$ is defined in terms of the likelihood model. (See Theorem 1, Corollary 1, and Example 1 in \cite{zahm2018certified}, and their proofs, for details.)

  Because $U$ is the matrix containing the first eigenvectors of $H$, we have our final result,
   \[\trace\left[ (I_d-U_rU_r^\top )H(I_d-U_rU_r^\top ) \right] = \lambda_{r+1}+\hdots+\lambda_d.\]

\subsection{Proof of Proposition \ref{prop:GreedyConvergence_condition}}\label{proof:GreedyConvergence_condition}

 We define \[R_\ell = \Dkl( (U_r^{\ell})^\top )_\sharp \pi_{\ell-1} || \rho_r ).\]
  Replacing $\pi$ by $\pi_{\ell-1}$ in \eqref{eq:KL_Tstar_Decomp} allows us to write $\Dkl(\pi_{\ell-1}||(T_\ell)_\sharp \rho) = \Dkl(\pi_{\ell-1}||\rho) - R_\ell$ so that 
  \[\Dkl(\pi_\ell|| \rho) = \Dkl(\pi||\rho) - \sum_{k=1}^{\ell-1} R_k.\]
 In particular $R_k$ converges to 0 and, because of \eqref{eq:GreedyConvergence_condition}, we have 
\[  \sup_{\substack{ U\in\R^{d\times d} \\ \text{s.t.} UU^\top =I_d}} \Dkl( (U_r^\top )_\sharp \pi_{\ell-1} || \rho_r ) \underset{\ell\rightarrow\infty}{\longrightarrow} 0.\]
 By Proposition 14.2 in \cite{huber1985projection}, $\pi_{\ell-1}$ converges weakly to $\rho$. Then $(T_1\circ\hdots\circ T_\ell)_\sharp\rho$ converges weakly to $\pi$.

 				
\section{Triangular maps}\label{app:triangular-maps}
One class of transport maps we consider in our numerical experiments (i.e., to approximate $\tau$ in \eqref{eq:LazyMap}, as a building block within the lazy structure) are lower triangular maps of the form, 
\begin{equation}
  \label{eq:kr-transport}
  T({\bf x}) = \left[
    \begin{array}{l}
      T_{1}(x_1)\\
      T_{2}(x_1, x_2)\\
      \quad \vdots \\
      T_{d}(x_1,\ldots,x_d)
    \end{array}
  \right]
\end{equation}
where each component $T_i$ is monotonically increasing with respect to $x_i$.
We will identify these transports with the set
$\mathcal{T}_> = \{
T:\mathbb{R}^d\rightarrow\mathbb{R}^d \, \vert \, T \text{ is
  triangular and } \partial_{x_i}T_i > 0
\}$. 
For any two distributions $\rho$ and $\pi$ on $\mathbb{R}^d$
that admit densities with respect to Lebesgue measure (also denoted by
$\rho$ and $\pi$, respectively) there exists a unique
transport $T\in\mathcal{T}_>$ such that $T_\sharp\rho = \pi$.
This transport is known as the Knothe--Rosenblatt (KR) rearrangement
\cite{knothe1957contributions,rosenblatt1952remarks,carlier2010knothe,bogachev2005triangular}.
Because $T$ is invertible,
the density of the \textit{pullback}
measure $T^\sharp\pi$ is given by
$T^\sharp\pi({\bm x}) = \pi \circ T({\bm x}) \det \nabla T({\bm x})$,
where $\det \nabla T({\bm x})$ is defined by
$\prod_{i=1}^d \partial_{x_i} T^{(i)}({\bm x}_{1:i})$.
We note here that $\det \nabla T({\bm x})$ is defined formally. Indeed, $T$ does
not need to be differentiable (in fact, $T$ inherits the same regularity as the densities of $\rho$ and $\pi$ \cite{bogachev2005triangular,santambrogio2015optimal}).
In \S\ref{sec:examples:elliptic}, and in the additional examples of
Appendix~\ref{app:numerical-algorithms}, we consider semi-parametric polynomial approximations to maps in $\mathcal{T}_>$. Specifically, we consider
the set $\mathcal{T}^\dagger_>\subset\mathcal{T}_>$ of maps
$T: ({\bf a}, {\bm x}) \mapsto T[{\bf a}]({\bm x})$ defined by
\begin{equation}
  \label{eq:monotone-component}
  T_{i}[{\bf c}_i,{\bf h}_i]({\bm x}) :=
  c[{\bf c}_i]({\bm x}_{1:i-1}) +
  \int_0^{x_i} \left( h[{\bf h}_i]({\bm x}_{1:i-1},t) \right)^2 dt \;,
\end{equation}
where ${\bf a}= \{ ({\bf c}_i,{\bf h}_i) \}_{i=1}^d$ denotes the coefficients of polynomials $c$ and $h$. As discussed in \S\ref{sec:LazyMaps}, we compute the transport map (i.e., an approximation to the KR rearrangement) between $\rho$ and $\pi$ as a
minimizer $T^\star$ of
\[\min_{T\in\mathcal{T}^\dagger_>} \dkl(T_\sharp \rho\Vert\pi) =
\min_{T\in\mathcal{T}^\dagger_>} \mathbb{E}_{\bm\rho}[\log \rho / T^\sharp \pi].\]
\cite{el2012bayesian,marzouk2016introduction,bigoni2016nips,spantini2018inference} provide more details and discussion.

 		
\section{Inverse auto-regressive flows}\label{app:iafmaps}
Another underlying class of transports that we use in our numerical experiments are inverse auto-regressive flows (IAFs). Introduced in \cite{Kingma2016}, IAFs are a class of normalizing flows parameterized using neural networks. IAFs are built as a composition of component-wise affine transformations, where the shift and scaling functions of each component only depend on earlier indexed variables. Each component of such a transformation can be expressed as
\[
T_i(x) = m_i(x_1, \dots x_{i-1}) + s_i(x_1, \dots x_{i-1}) x_i
\]
where the functions $m_i$ and $s_i$ are defined by neural networks. These maps are naturally lower triangular, and the Jacobian determinant is given by the product of the scaling functions of each component, i.e.,
\[
\det(\nabla T) = \prod_{i=1}^d s_i(x),
\]
allowing for efficient computation. Flows are typically comprised of several IAF stages with the components either randomly permuted or, as we choose, reversed in between each stage. For the results of \S\ref{sec:examples:blr} and \S\ref{sec:examples:bnn} we construct IAFs using $4$ stacked IAF layers. The autoregressive networks each use $2$ hidden layers, a hidden dimension of $128$ and ELU activation functions. Each map was trained using Adam \cite{kingma15adam} with step size $10^{-3}$ for $20000$ iterations. The optimization objective (i.e., the ELBO) was approximated using $100$ independent samples from $\rho$ at each iteration. 			

\section{Generalized linear models and lazy structure}
\label{app:glms}
Here we discuss how generalized linear models 
may naturally admit lazy structure. We consider a Bayesian logistic regression problem as an example, but the same result follows for other generalized linear models. Let $M$ denote the number of observations in a data set and $N$ denote the number of covariates or features. In \S\ref{sec:examples:blr}, we considered $N = 500$ covariates. The low rank problem used $M = 20$ observations and the full rank problem used $M = 605$ observations.  For each observation $i = 1,\dots,M$ and covariate $j = 1,\dots,N$, we denote the observed covariates by $f_{ij} \in \R$, the observations as $y_i \in \{0,1\}$, and the model parameters as $x_j \in \R$. The single observation likelihood is then defined as
\[
\ell_i(\bfx) = P(\bfx,\bff_i)^{y_i}(1 - P(\bfx,\bff_i))^{1 - y_i}
\]
where the quantity
\[
P(\bfx,\bff_i) = \left(1 + \exp( - \bfx^T \bff_i) \right)^{-1} = \text{sigmoid}( \bfx^T \bff_i) 
\]
models the probability that $y_i = 1$. This has the form of a generalized linear model, i.e., the likelihood depends on a linear function of the covariates, $\bfx^T \bff_i$.
The gradient of the log likelihood then has the form
\[\nabla_\bfx \log(\ell_i(\bfx)) = \bff_i h(\bfx;\bff_i,y_i)\]
for some function $h$. Assuming independence of the observations, the likelihood of the data set can be written as
\[\nabla_\bfx \log(\sfL(\bfx)) = \sum_{i = 1}^M \bff_i h(\bfx;\bff_i,y_i) = \bfF \bfh(\bfx).\]
We can then express the diagnostic matrix $H$ as
\[H = \int \left(\nabla \log(\sfL(\bfx)) \right) \left(\nabla \log(\sfL(\bfx)) \right)^T d \pi = \bfF \left[ \int \left(\bfh(\bfx) \right) \left(\bfh(\bfx) \right)^T d \pi \right] \bfF^T,\]
and so the rank of $H$ is bounded by the rank of the feature matrix $\bfF$ which is at most $\min(N,M)$. If $M<N$, we are in the exactly lazy setting, where $r = M$. We also note that  $\bfF$ may be low rank due to redundancy in the measurements, meaning when $\bff_i$ is nearly aligned with $\bff_j$; more generally, it might exhibit some spectral decay.

 					
\section{The use of $H^{\text{B}}$ vs $H$}\label{app:biasofH}
\begin{wrapfigure}{R}{0.43\columnwidth}
	\centering
	\vspace{-3ex}
	\includegraphics[width=6cm]{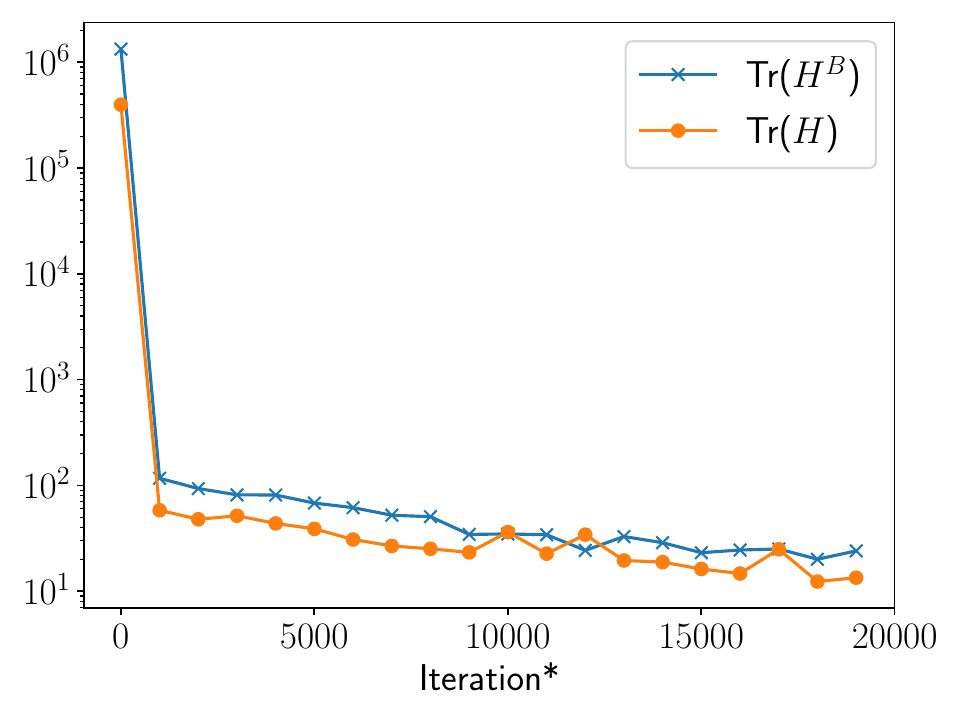}
	\caption{The two trace diagnostics through out the training of $U_r$-IAF on the low rank Bayesian logistic regression problem . \vspace{-1em}}
	\label{fig:blr:lr:comparetraces}
\end{wrapfigure}
We note in \S \ref{sec:LazyMaps} that a practical implementation of Algorithm \ref{alg:LazyMap} requires the numerical approximation of the diagnostic matrix $H$ defined by
\[
H = \int  \left ( \nabla\log\frac{\pi}{\rho} \right ) \left
  (\nabla\log\frac{\pi}{\rho} \right )^\top\d\pi.
\]

This poses a challenge as we cannot generate samples from $\pi$. We can obtain an (asymptotically) unbiased estimate of $H$ using self-normalized important sampling (IS), but as we comment in the main text, this estimate typically has large variance when the IS instrumental/biasing distribution is far from $\pi$. Instead, we can use the diagnostic matrix $H^{\text{B}}$, where the expectation is instead taken with respect to the reference density $\rho$
\[
H^{\text{B}} = \int \left (\nabla\log\frac{\pi}{\rho} \right ) \left (\nabla\log\frac{\pi}{\rho} \right )^\top\d\rho.
\]
Unbiased estimates of $H^{\text{B}}$ can be computed easily using direct Monte Carlo sampling, but these are of course biased estimates of $H$ in general. In this section we comment on the use of this biased estimate in the error bound on the KL divergence, and find that this bias leads to a more conservative diagnostic. 

\begin{figure}
	\centering
	\begin{subfigure}[b]{.49\textwidth}
		\includegraphics[width=\textwidth]{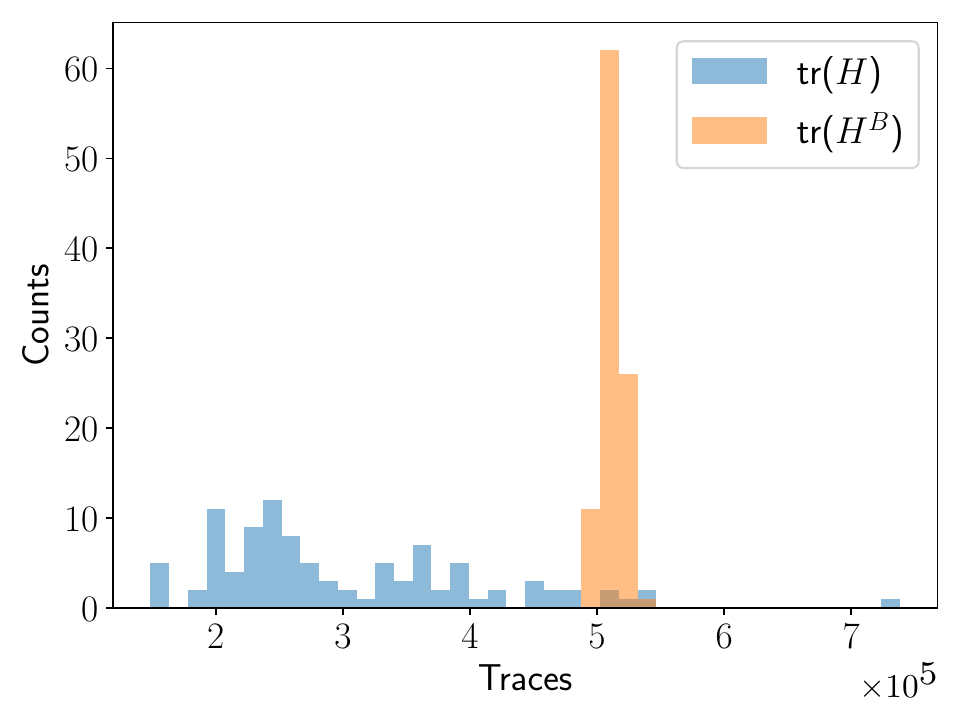}
		\caption{Before training}
		\label{fig:biasH:before}
	\end{subfigure}
	\hspace{2pt}
	\begin{subfigure}[b]{.49\textwidth}
		\includegraphics[width=\textwidth]{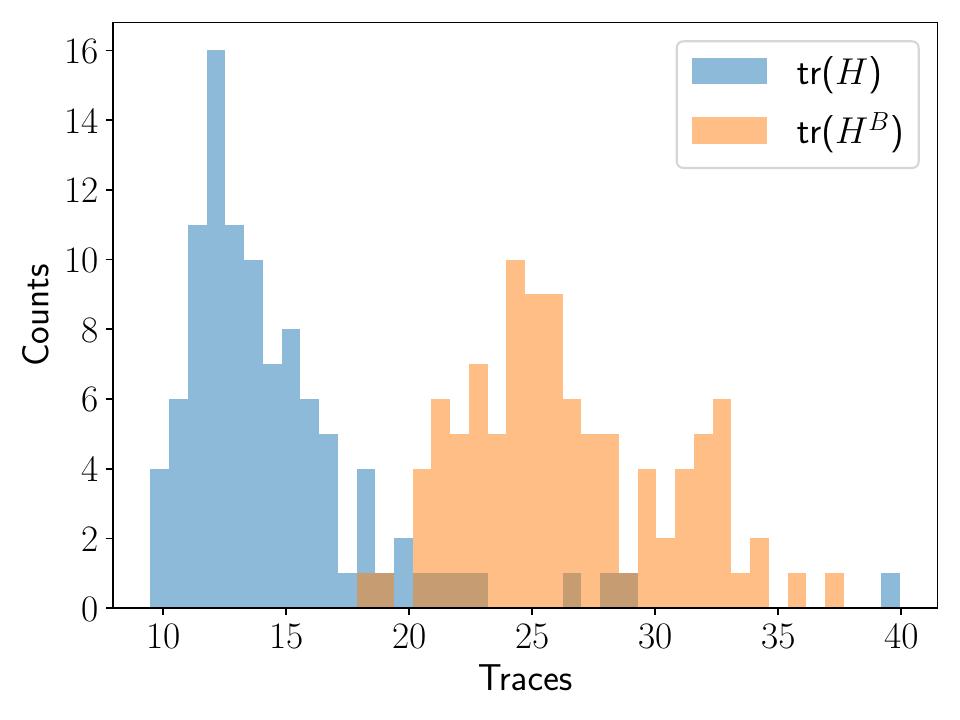}
		\caption{After training}
		\label{fig:biasH:after}
	\end{subfigure}
	
	\caption{
		Histograms of $\trace(H^{\text{B}})$ and $\trace(H)$ before and after training for the low rank logistic regression problem.
	}
	\label{fig:app:biasofH}
\end{figure}

Figure \ref{fig:biasH:before} shows histograms of $100$ estimates of $\trace(\widehat{H})$ (where $\widehat{H}$ is a self-normalized IS estimate of $H$) and $\trace(\widehat{H}^\text{B})$ (where $\widehat{H}^\text{B}$ is a Monte Carlo estimate of $H^{\text{B}}$) for the low-rank logistic regression problem of \S \ref{sec:examples:blr}. Each estimate was constructed from $K=500$ samples. We see that the variance of $\trace(\widehat{H})$ is higher than that of $\trace(\widehat{H}^\text{B})$. Figure \ref{fig:biasH:after} shows similar histograms for $\trace(\widehat{H}_\ell)$ and $\trace(\widehat{H}^\text{B}_\ell)$ after the training of the transport map. We see that the bias has decreased now that the approximate posterior is close to the true posterior; where indeed $H^{\text{B}}_\ell$ is closer to $H_\ell$.  The variance of the IS estimate $\trace(\widehat{H}_\ell)$ has decreased significantly as well. Figure \ref{fig:blr:lr:comparetraces} shows the two trace diagnostics computed throughout the training of the $U_r$-IAF lazy map. We see that $\frac12 \trace(H^{\text{B}}_\ell) > \frac12 \trace(H_\ell)$ throughout the training process, meaning it is a more conservative error bound for this particular problem.


\section{Numerical algorithms}\label{app:numerical-algorithms}

Here we describe the numerical algorithms required by the lazy map framework. Algorithm \ref{alg:compute-H-rho} assembles the numerical estimate $\widehat{H}^\text{B}$ via some quadrature rule (e.g. Monte Carlo, Gauss quadrature \cite{Golub1969}, sparse grids \cite{Smolyak1963}, ect.) of
$H^\text{B}=\int (\nabla\log\frac{\pi}{\rho} )(\nabla\log\frac{\pi}{\rho} )^\top\d\rho$.

Algorithm \ref{alg:compute-subspace} computes the eigenvectors $U$
satisfying Proposition \ref{prop:LogSobResult} and discerns between the
subspace of relevant directions $\Span(U_r)$ and its orthogonal complement
$\Span(U_\perp)$.

Algorithm \ref{alg:compute-map} outlines the numerical solution of the
variational problem
\begin{equation}
  \label{eq:app:alg:reverse-kl-minimization}
  T[{\bf a}^\star] = \arg\min_{T[{\bf a}] \in \Tc} \Dkl\left(T[{\bf a}]_\sharp \rho \middle\Vert \pi \right).
\end{equation}
For the sake of simplicity we fix the complexity the underlying transport class $\Tc$ and the sample
size $m$ used in the discretization of the KL divergence.
Alternatively one could adaptively increase the complexity and the
sample size to match a prescribed tolerance,
following the procedure described in \cite{bigoni2016nips}.
For the examples presented in this work,
the variational problem is solved either with the Adam optimizer \cite{kingma15adam} or with the
Broyden–Fletcher–Goldfarb–Shanno (BFGS) quasi-Newton
method \cite{Broyden1970}.
One could switch to a full Newton method
if the Hessian of $\pi$ or its action on a vector are available.

Algorithms \ref{alg:lazy-map-construction} and \ref{alg:compute-deep-map}
are numerical counterparts of
Algorithms \ref{alg:LazyMap} (constructing a lazy map) and \ref{alg:LayersOfLazyMaps} (constructing a deeply lazy map) respectively.

\begin{algorithm}
  \caption{
    Given the quadrature rule $(x_i, w_i)_{i=1}^m$ 
    with respect to the base distribution $\rho$, 
    and the unnormalized density $\pi$, compute an approximation
    to $H^\text{B}=\int (\nabla\log\frac{\pi}{\rho} )(\nabla\log\frac{\pi}{\rho} )^\top\d\rho$.
  }
  \begin{algorithmic}[1]
    \Procedure{ComputeH}{
      $(x_i, w_i)_{i=1}^m$, 
      $\pi$
    }
    \State Assemble
    $$
    \widehat{H}^\text{B}= \sum_{i=1}^m
    \left(\nabla_{\bf x} \log \frac{\pi(x_i)}{\rho(x_i)} \right)
    \left(\nabla_{\bf x} \log \frac{\pi(x_i)}{\rho(x_i)} \right)^T w_i
    $$
    \Return $\widehat{H}^\text{B}$
    \EndProcedure
  \end{algorithmic}
  \label{alg:compute-H-rho}
\end{algorithm}

\begin{algorithm}
  \caption{
    Given the matrix
    $\widehat{H}^\text{B} \approx H^\text{B}$, the tolerance $\varepsilon$, and a maximum lazy rank $r_\text{max}$,
    find the matrix $U \coloneqq [ U_r \, \vert \, U_\perp ]$
    that satisfies Proposition \ref{prop:LogSobResult}.
  }
  \begin{spacing}{1.3}
  \begin{algorithmic}[1]
    \Procedure{ComputeSubspace}{
      $\widehat{H}^\text{B}$, $\varepsilon$, $r_{{\text{max}}}$
    }
    \State Solve the eigenvalue problem $\widehat{H}^\text{B} X =
    \Lambda X$
    \State Let $r= r_{{\text{max}}} \wedge \min \{ r\leq d ~: \frac{1}{2}\sum_{i>r}\lambda_{i} \leq \varepsilon \}$
    \State Define $U_r = [X_{:,1},\ldots,X_{:,r}]$ and $U_\perp = [X_{:,r+1},\ldots,X_{:,n}]$
    \State\Return $U_r$, $U_\perp$, $r$
    \EndProcedure
  \end{algorithmic}
  \end{spacing}
  \label{alg:compute-subspace}
\end{algorithm}

\begin{algorithm}
  \caption{
    Given the quadrature rule $(x_i, w_i)_{i=1}^m$ 
    with respect to the base distribution $\rho$,
    the unnormalized target density $\pi$,
    a set of underlying class of transport maps $\mathcal{T}$,
    a tolerance $\varepsilon_{\text{map}}$,
    find the optimal map parameters ${\bf a}^\star$
    such that $T[{\bf a}]_\sharp \rho \propto \pi$ by minimizing
    \eqref{eq:app:alg:reverse-kl-minimization}.
  }
  \begin{algorithmic}[1]
    \Procedure{ComputeMap}{
      $(x_i, w_i)_{i=1}^m$,
      $\pi$, 
      $\mathcal{T}_{\bf a}$,
      $\varepsilon_{\text{map}}$
    }
    \State Solve (e.g., via a stochastic or deterministic optimization method),
    \begin{equation*}
      \begin{aligned}
        T[{\bf a}^\star] &= \argmin_{T[{\bf a}]\in\mathcal{T}} 
        \underbrace{-\sum_{i = 1}^m \log(T[{\bf a}]^\sharp \pi(x_i)) w_i}_{\Jc[{\bf a}]} \;, \\
        & \text{based on some stopping criteria, e.g.,}\quad \Vert\nabla_{\bf a}\Jc[{\bf a}^\star]\Vert_2 < \varepsilon_{\text{map}}
      \end{aligned}
    \end{equation*}
    
    \State \Return $T[{\bf a}^\star]$
    \EndProcedure
  \end{algorithmic}
  \label{alg:compute-map}
\end{algorithm}

\begin{algorithm}
  \caption{
    Given the quadrature rule $(x_i, w_i)_{i=1}^m$ 
    with respect to the base distribution $\rho$, the unnormalized density $\pi$,
    the matrix
    $\widehat{H}^\text{B}\approx H^\text{B}$,
    the rank truncation tolerance $\varepsilon_{\text{r}}$,
    the maximum lazy rank $r_{\text{max}}$,
    the class of transport maps
    $\mathcal{T}$ and the target tolerance
    $\varepsilon_{\text{map}}$
    for learning the map $\tau$, identify the optimal lazy
    map $T$.
  }
  \begin{spacing}{1.3}
  \begin{algorithmic}[1]
    \Procedure{LazyMapConstruction}{
      $(x_i, w_i)_{i=1}^m$,
      $\pi$,
      $\widehat{H}^\text{B}$,
      $\varepsilon_{\text{r}}$,
      $r_{{\text{max}}}$,
      $\mathcal{T}$,
      $\varepsilon_{\text{map}}$,
    }
    \State $U_r, U_\perp, r \gets$ 
    \Call{ComputeSubspace}{
      $\widehat{H}^\text{B}$, $\varepsilon_{\text{r}}$, $r_{\text{max}}$
    } \Comment{Algorithm \ref{alg:compute-subspace}}

    \State Define
    $\hat{\pi}(x)\coloneqq (U_r\vert U_\perp)^\sharp\pi(x) = \pi\circ (U_r\vert U_\perp)\, x$


    \State Build the quadrature $(x_i, w_i)_{i=1}^{m}$
    with respect to $\mathcal{N}(0,I_d)$

    \State Define
    $ \Tc_{r} = \left\{
    T[{\bf a}](z) = \left[
        \tau[{\bf a}]( z_1,\hdots,z_r)^\top, z_{r+1}, \cdots, z_d
      \right]^\top
      \;\middle\vert\;
      \tau[{\bf a}] \in \Tc
    \right\}$
    
    \State $T[{\bf a}^\star] \gets$
    \Call{ComputeMap}{
      $(x_i, w_i)_{i=1}^{m}$,
      $\hat{\pi}$,
      $\mathcal{T}_{r}$,
      $\varepsilon_{\text{map}}$
    } \Comment{Algorithm \ref{alg:compute-map}}

    \State Define $L(z) \coloneqq (U_r\vert U_\perp) \circ T[{\bf a}](z)$
    

    \State \Return $L$
    \EndProcedure
  \end{algorithmic}
  \end{spacing}
  \label{alg:lazy-map-construction}
\end{algorithm}

\begin{algorithm}
  \caption{
    Given the quadrature rule $(x_i, w_i)_{i=1}^m$ 
    with respect to the base distribution $\rho$, the target density $\pi$, a stopping tolerance $\varepsilon$
    and a maximum number of lazy layers $\ell_{\text{max}}$,
    compute a deeply lazy map.
    See Algorithm \ref{alg:lazy-map-construction} for the definition
    of the remaining arguments.
  }
  \begin{spacing}{1.3}
    \begin{algorithmic}[1]
      \Procedure{LayersOfLazyMapsConstruction}{$(x_i, w_i)_{i=1}^m$, $\pi$, $\varepsilon$, $r$, $\ell_{{\text{max}}}$, $\Tc$, $\varepsilon_{\text{map}}$}
      \State Set $\pi_0=\pi$ and $\ell=0$
      \State Build the quadrature $(x_i, w_i)_{i=1}^{m}$ with respect to $\mathcal{N}(0,I_d)$
      \State Compute $\widehat{H}^\text{B}_\ell =$ \Call{ComputeH}{
        $(x_i, w_i)_{i=1}^m$, $\pi_\ell$
      }
      \While{$\ell \leq \ell_{{\text{max}}}$ and $\frac{1}{2}\trace(\widehat{H}^\text{B}_\ell)\geq \varepsilon$}
      \State $\ell\gets\ell+1$
      \State $T_\ell \gets $ \Call{LazyMapConstruction}{
        $(x_i, w_i)_{i=1}^m$, $\pi_{\ell-1}$, $\widehat{H}^\text{B}_\ell$, $0$, $r$, $\Tc$, $\varepsilon_{\text{map}}$
      } \Comment{Algorithm \ref{alg:lazy-map-construction}} 
      \State Update $\frak{T}_{\ell}=\frak{T}_{\ell-1}\circ T_\ell$ 
      \State Compute $\pi_\ell = (\frak{T}_{\ell})^\sharp \pi$
      \State Build the quadrature $(x_i, w_i)_{i=1}^{m}$ with respect to $\mathcal{N}(0,I_d)$
      \State Compute $\widehat{H}^\text{B}_\ell =$ \Call{ComputeH}{
        $(x_i, w_i)_{i=1}^m$, $\pi_\ell$
      }
      \EndWhile
      
      \State\Return $\frak{T}_\ell=T_1\circ\cdots\circ T_\ell$
      \EndProcedure
    \end{algorithmic}
  \end{spacing}
  \label{alg:compute-deep-map}
\end{algorithm}

	
\newpage 

\section{Numerical examples: additional details and experiments}

In this section, we provide more details concerning our numerical examples and present several other numerical experiments.

\subsection{Additional details: Bayesian logistic regression}\label{app:example:blr}
Here we provide addition details and results for the Bayesian logistic regression problems discussed in \S \ref{sec:examples:blr}. We begin by further describing the UCI Parkinson's disease data set \cite{parkinsonsUCI}. The $500$ features we consider consist of the patient sex, and audio extensions from a patient recording. The data set includes data from $3$ independent recordings from $188$ Parkinson's disease patients and a control group of $64$ individuals, totaling $756$ observations in all. The low rank problem considers $20$ observations where we use observations from $20$ different individuals.

We imposed a non-informative prior of $\sfN(0,10^2 I_d)$ on the parameters. Samples from the prior can be transformed to match those of a standard normal distribution via a \emph{whitening} transformation, i.e.
\[
z \sim \sfN(0,10^2 I_d), \  Wz:= \frac{1}{10} z \sim \sfN(0,I_d),
\]
where we let $W$ denote this whitening operation. We consider the transformed posterior
\[
\widetilde{\pi}(x) \propto \sfL(W^{-1} x) \rho(x)
\]
where the prior has been replaced with a standard normal distribution. This whitened posterior relates to the true posterior by $\widetilde{\pi} = W_\sharp \pi$. We see that solving this transformed problem is equivalent to solving the original, and that working with this whitened problem directly exposes lazy structure by matching the form of \ref{eq:LazyMap_VS_Ridge}. A similar whitening process is followed for each of the numerical experiments.

\begin{figure}
	\centering
	\begin{subfigure}[b]{.49\textwidth}
		\includegraphics[width=\textwidth]{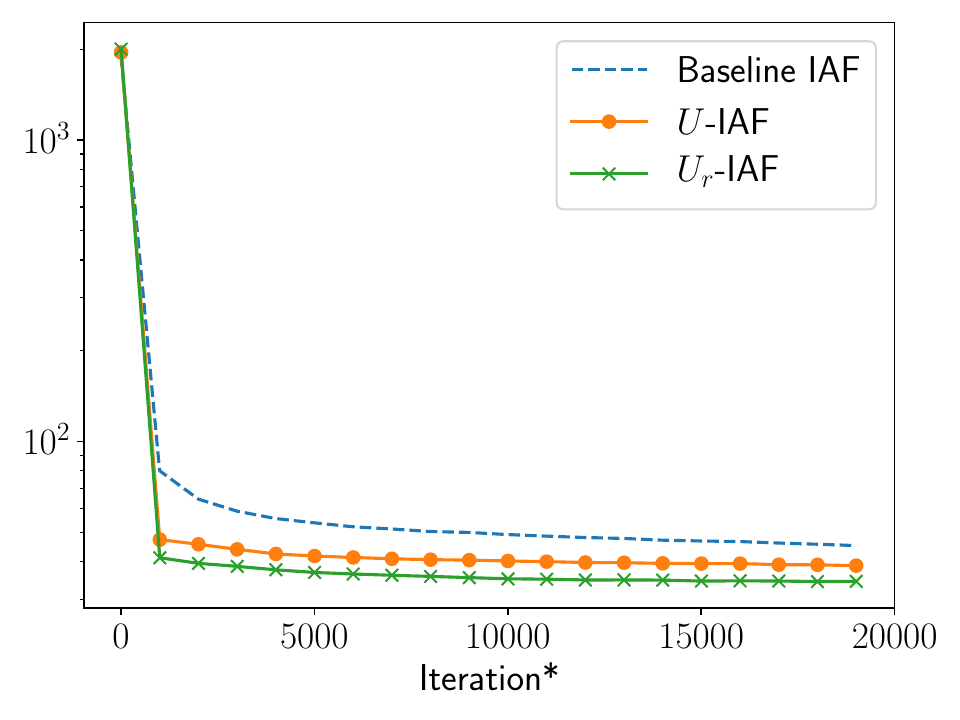}
		\caption{Low rank - Negative ELBO}
		\label{fig:blr:lr:elbo}
	\end{subfigure}
	\hspace{2pt}
		\begin{subfigure}[b]{.49\textwidth}
		\includegraphics[width=\textwidth]{convergence-plots/blr_lr_elbos.pdf}
		\caption{Full rank - Negative ELBO}
		\label{fig:blr:fr:elbo}
	\end{subfigure}

	\begin{subfigure}[b]{.49\textwidth}
	\includegraphics[width=\textwidth]{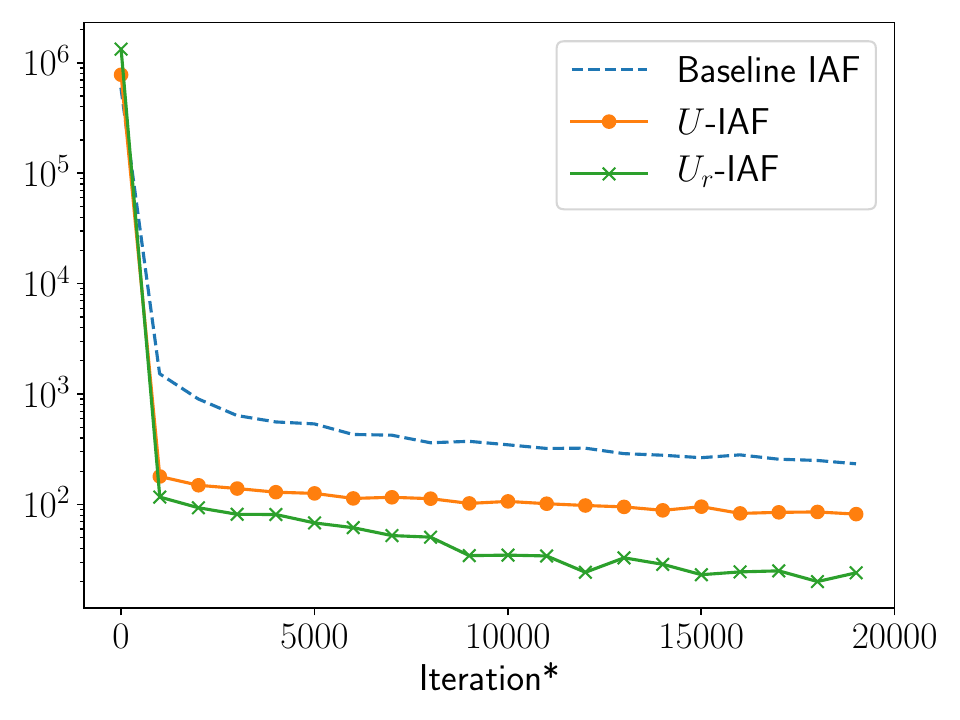}
	\caption{Low rank - $\trace(H^{\text{B}})$}
	\label{fig:blr:lr:traces}
\end{subfigure}
	\hspace{2pt}
		\begin{subfigure}[b]{.49\textwidth}
		\includegraphics[width=\textwidth]{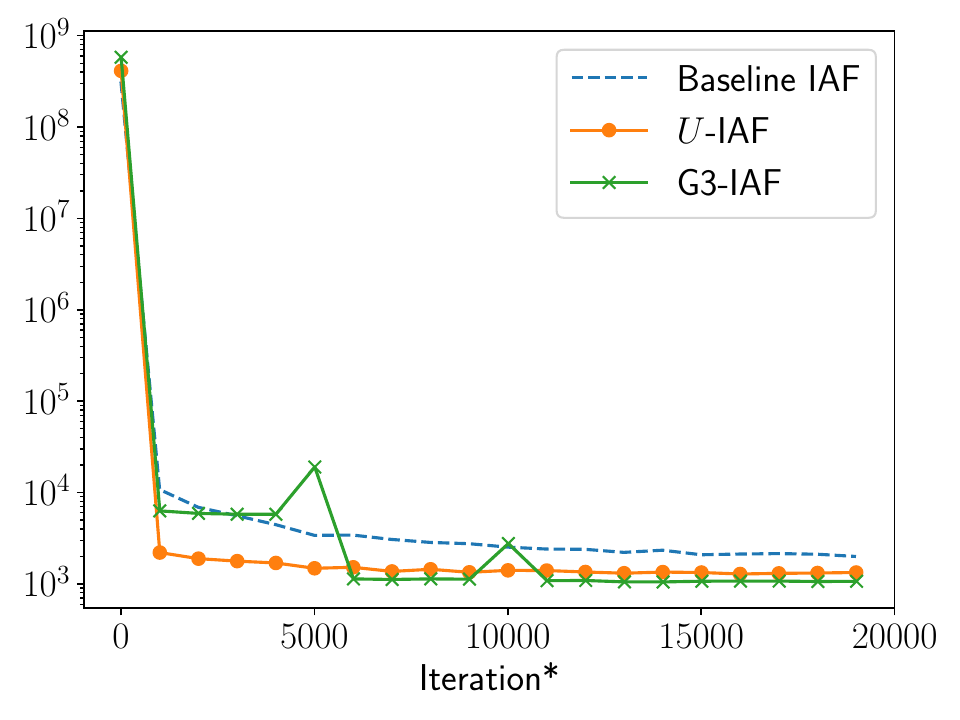}
		\caption{Full rank - $\trace(H^{\text{B}})$}
		\label{fig:blr:fr:traces}
	\end{subfigure}

	\begin{subfigure}[b]{.49\textwidth}
	\includegraphics[width=\textwidth]{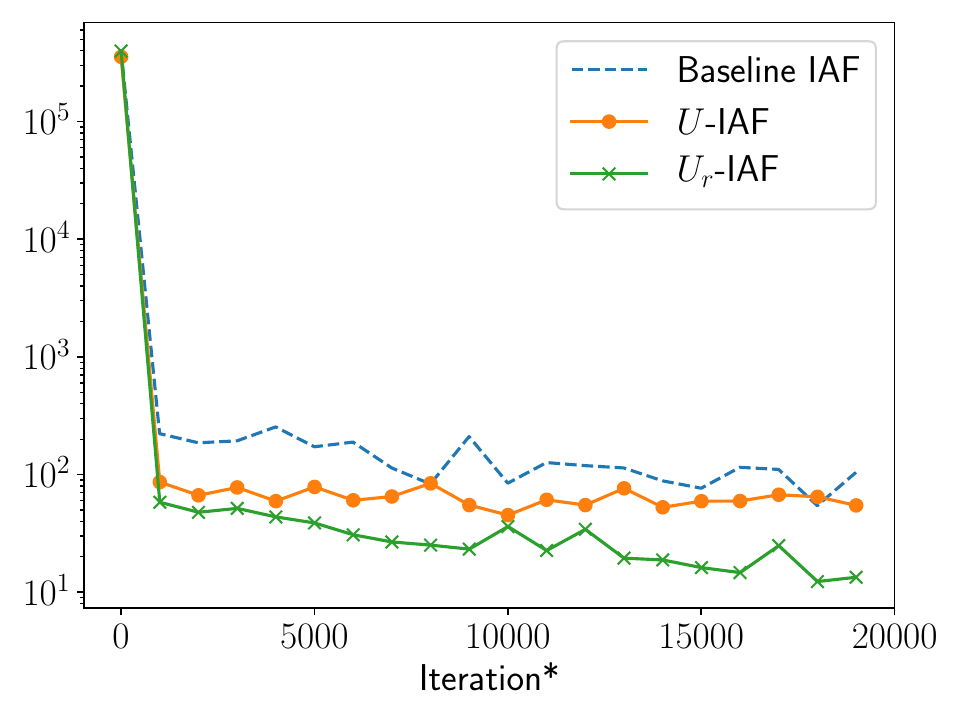}
	\caption{Low rank - $\trace(H)$}
	\label{fig:blr:lr:is-traces}
\end{subfigure}
\hspace{2pt}
	\begin{subfigure}[b]{.49\textwidth}
	\includegraphics[width=\textwidth]{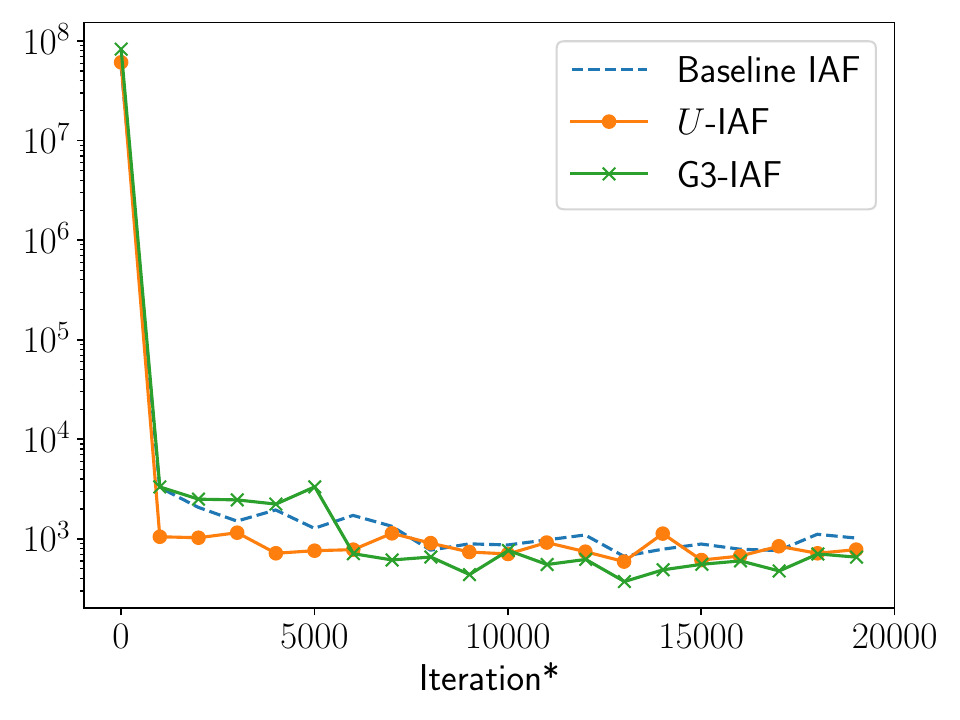}
	\caption{Full rank - $\trace(H)$}
	\label{fig:blr:fr:is-traces}
\end{subfigure}
	\caption{
		Mean training plots for the low rank (left) and full rank (right) Bayesian logistic regression problems across $10$ optimization runs. $^*$The $x$-axes include the cost of forming the matrices $H_\ell$ to determine the subspace $U^{\ell}$ in terms of gradient evaluations. Each matrix is computed using $500$ gradient evaluations, the same cost as $5$ optimization steps.
	}
	\label{fig:app:blr:lr}
\end{figure}

Figure \ref{fig:app:blr:lr} shows mean performance metrics through out the training process for each of the maps considered. Each metric is computed with $500$ independent samples. For G3-IAF, the three lazy layers were trained for $5000$, $5000$ and $10000$ iterations, which can be seen as sharp decreases in the negative ELBO and trace diagnostics occur. In general we see faster convergence in terms of the number of iterations for maps using the lazy framework compared to the baselines.

\subsection{Additional details: Bayesian neural network}\label{app:example:bnn}

In \S \ref{sec:examples:bnn} we considered a Bayesian neural network that is also used as a test problem in \cite{liu2016stein} and \cite{Detommaso2018}. Bayesian neural networks generate high dimension inference problems, where the parameter dimension is the number of parameters in the underlying neural network. We considered the UCI yacht hydrodynamics data set \cite{yachtrockUCI}. In our example, the parameter dimension is $581$, given an input dimension of $6$, one hidden layer of dimension $20$, and output layer of dimension $1$. We use sigmoid activation functions in the input and hidden layer, and a linear output layer. The prior on the model parameters is taken to be zero mean Gaussian with a variance of $100$.

\begin{figure}
	\centering
	\begin{subfigure}[b]{.49\textwidth}
		\includegraphics[width=\textwidth]{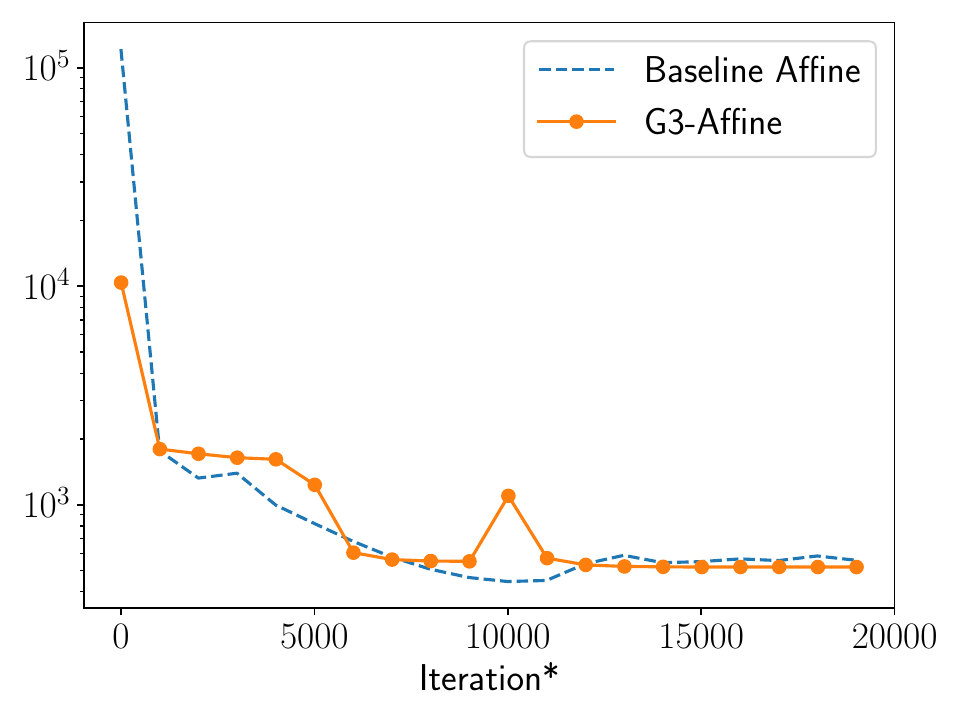}
		\caption{Negative ELBO}
		\label{fig:bnn:elbo}
	\end{subfigure}
	\hspace{2pt}
	\begin{subfigure}[b]{.49\textwidth}
		\includegraphics[width=\textwidth]{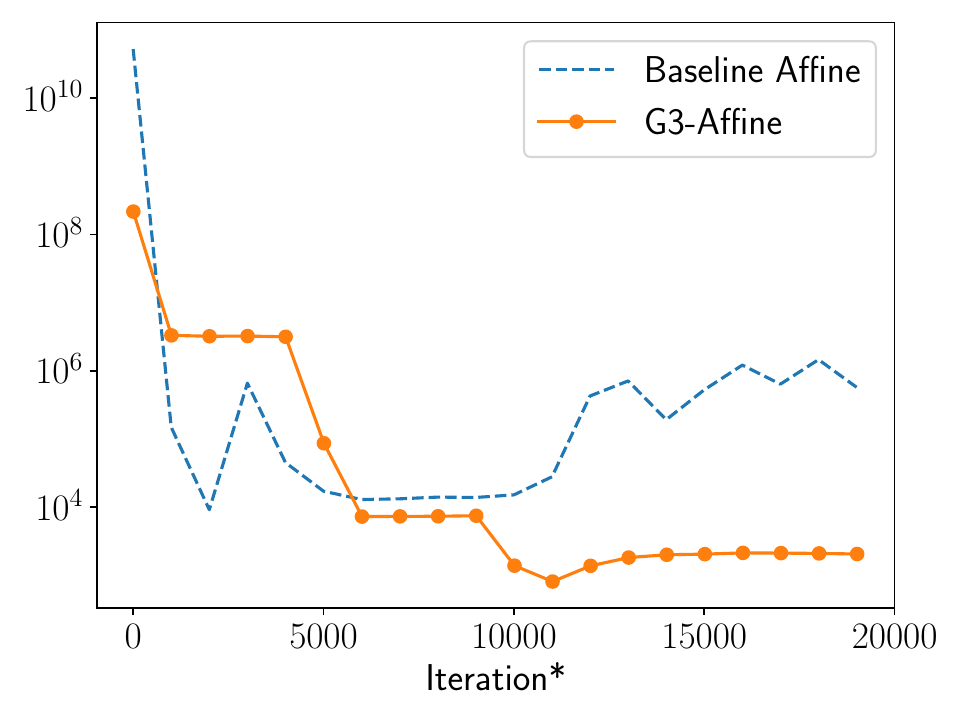}
	\caption{$\trace(H^{\text{B}})$}
		\label{fig:bnn:traces}
	\end{subfigure}
	\hspace{2pt}
	\begin{subfigure}[b]{.49\textwidth}
		\includegraphics[width=\textwidth]{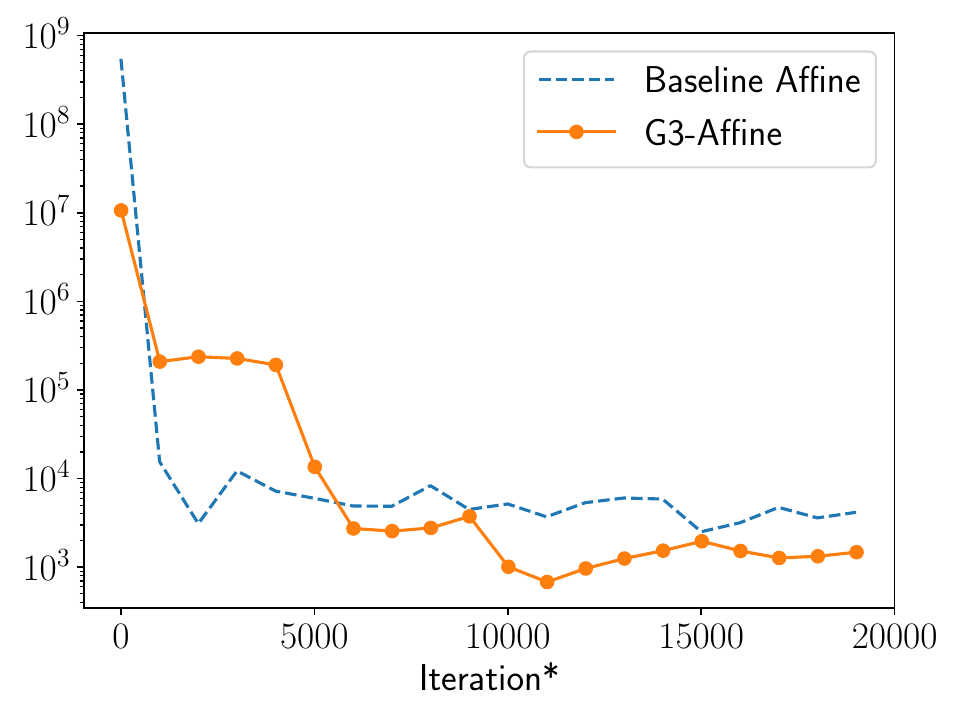}
	\caption{$\trace(H)$}
		\label{fig:bnn:is-traces}
	\end{subfigure}
	\hspace{2pt}
	\begin{subfigure}[b]{.49\textwidth}
		\includegraphics[width=\textwidth]{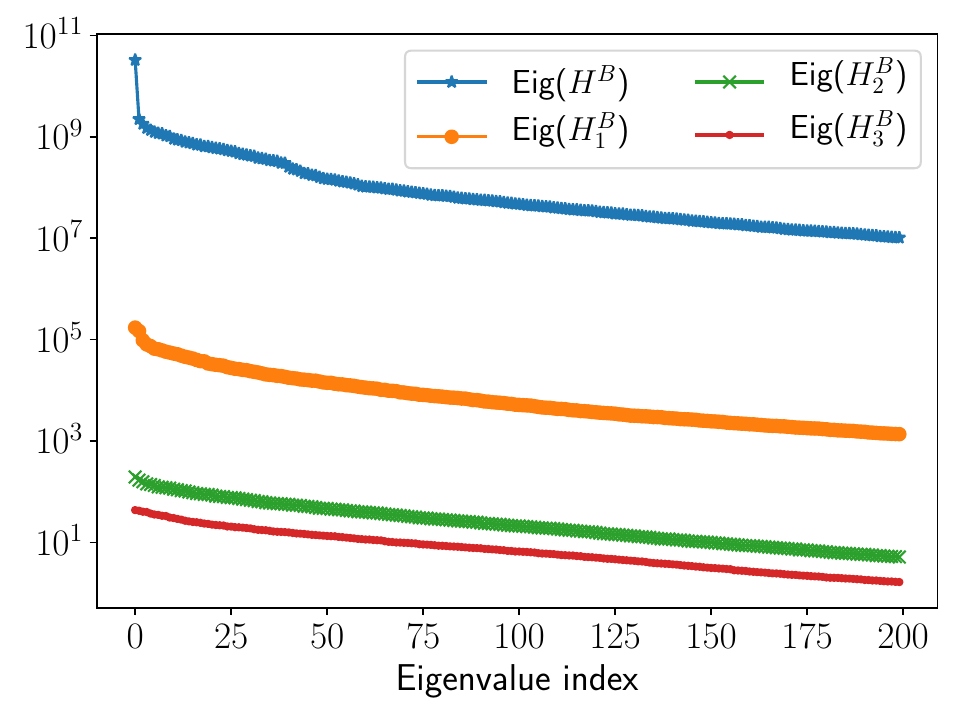}
		\caption{Spectrum of $H^\text{B}_\ell$ before and after training each lazy layer.}
		\label{fig:bnn:eigvalues}
	\end{subfigure}
	\caption{
		(\subref{fig:bnn:elbo},\subref{fig:bnn:traces},\subref{fig:bnn:is-traces}) Mean training plots for the Bayesian neural network problem across $10$ optimization runs. $^*$The $x$-axes include the cost of forming the matrices $H_\ell$ to determine the subspace $U^{\ell}$ in terms of gradient evaluations. Each matrix is computed using $581$ gradient evaluations, approximately the same cost as $6$ optimization steps. (\subref{fig:bnn:eigvalues}) Analogous plot to Figure \ref{fig:blr_fullrank_eigenvalues} for the Bayesian neural network problem. The spectrum of the diagnostic matrices $H^{\text{B}}_\ell$ flatten and fall as the approximation to the posterior improves with each lazy layer.
	}
	\label{fig:app:bnn}
\end{figure}

Here we consider affine maps, i.e., maps of the form $T(x) = \mu + Lx$, where $L \in \R^{d \times d}$ denotes a lower triangular matrix and $\mu \in \R^d$ a constant vector. The approximate posteriors in this case are indeed Gaussian distributions with mean $\mu$ and covariance $\Sigma = L L^T$. We note that the final approximate posterior given by the G3-affine transport map is also Gaussian given that the composition of affine functions is affine. Therefore the performance benefits we see may come from avoiding sub-optimal minima of the KL divergence. We see stabler training in terms of the performance metrics in Figure \ref{fig:app:bnn}. For G3-affine, layers were trained for $5000$, $5000$ and $10000$ iterations, where we see sharp decreases in each of the diagnostics.
\subsection{Additional details: High-dimensional elliptic PDE inverse problem}\label{app:examples:elliptic}

\begin{figure}
  \centering
  \begin{subfigure}[b]{.29\textwidth}
    \includegraphics[width=\textwidth]{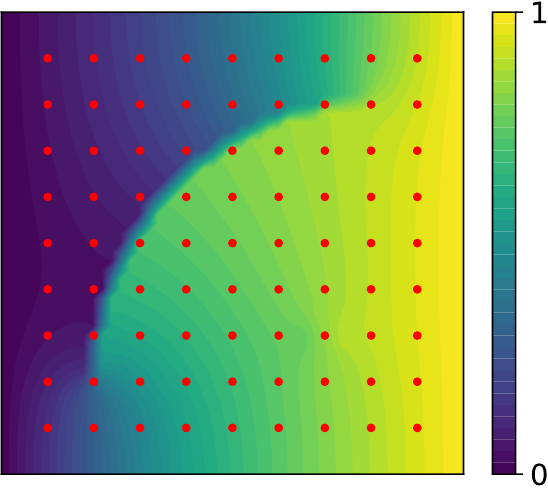}
    \caption{Solution $\Sc(\kappab^\star)$}
    \label{fig:ex:elliptic:true-solution}
  \end{subfigure}
  \hspace{2pt}
  \begin{subfigure}[b]{.31\textwidth}
    \includegraphics[width=\textwidth]{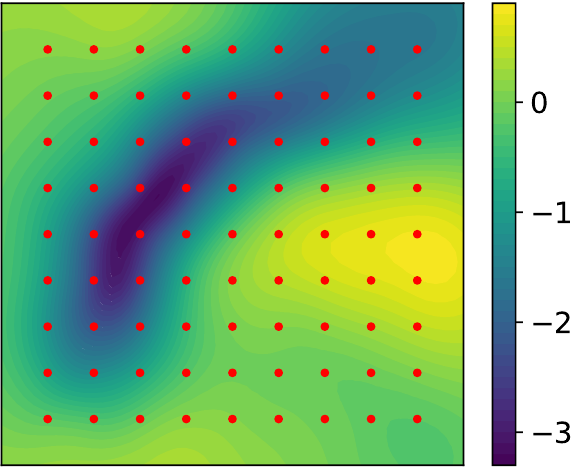}
    \caption{$\mathbb{E}[\kappab\vert\yb^\star]$}
    \label{fig:ex:elliptic:mean-log-kappa}
  \end{subfigure}
  \hspace{2pt}
  \begin{subfigure}[b]{.32\textwidth}
    \includegraphics[width=\textwidth]{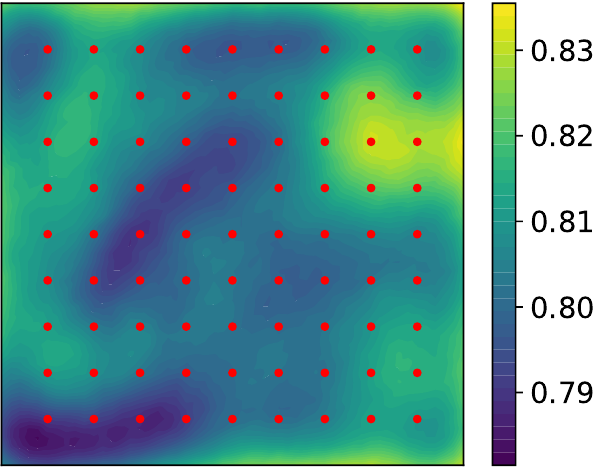}
    \caption{$\mathrm{Std}[\kappab\vert\yb^\star]$}
    \label{fig:ex:elliptic:std-log-kappa}
  \end{subfigure}
  \caption{
    Additional figures for the elliptic problem with
    unknown diffusion coefficient.
    Figure
    (\subref{fig:ex:elliptic:true-solution}) shows the solution $u$
    corresponding to the field in Figure \ref{fig:ex:elliptic:true-kappa}.
    Figures
    (\subref{fig:ex:elliptic:mean-log-kappa}) and
    (\subref{fig:ex:elliptic:std-log-kappa})
    show the mean and the standard deviation of the posterior distribution.
  }
  \label{fig:app:elliptic}
\end{figure}

Here we explain how the numerical discretization of the PDE enters
the Bayesian inference formulation.
We denote by $\Sc$ the map $\kappab\mapsto u$, mapping the discretized coefficient
to the numerical solution of equation \ref{eq:elliptic}.
The observation map is defined by the operator
$B_i(u)\coloneqq\int_\Dc u\,\phi_i\,\d\xb$, where
$\phi_i(\xb)\coloneqq\exp[-\Vert\sbb_i-\xb\Vert_2^2/(2\delta^2)]/\gamma_i$,
$\{\sbb_i\}_{i=1}^n\in\Dc$ are observation locations, $\delta=10^{-4}$,
and $\gamma_i$ are normalization constants so that $\int_\Dc \phi_i\,\d\xb=1$ for all $i=1,\ldots,n$.
The parameter-to-observation map is then defined by
$\Fc:\kappab\mapsto [B_1(\Sc(\kappa)), \ldots,B_n(\Sc(\kappa))]^\top$.
The coefficient $\kappa$ is endowed with the distribution
$\kappa\sim\Nc(0,\Cc(\xb,\xb^\prime))$, where
$\Cc(\xb,\xb^\prime)\coloneqq\exp(-\Vert \xb -\xb^\prime\Vert_2)$
is the Ornstein–Uhlenbeck (exponential) covariance kernel.
Letting $\Sigma$ be the discretization of $\Cc$ over the finite element mesh,
we define the likelihood to be
$\Lc_{\yb}(\zb)\propto\exp\left(-\left\Vert \yb - \Fc(\Sigma^{1/2}\zb) \right\Vert_{\Sigma_{\text{obs}}^{-1}}\right)$.
We stress here that the model is computationally demanding:
the evaluation of $\pi(\zb)$ and $\nabla\pi(\zb)$ require approximately
$1$ second.

Figure \ref{fig:app:elliptic} shows the observation generating solution
$u^\star = \Sc(\kappab)$,
the posterior mean $\mathbb{E}[\kappab\vert\yb^\star]$ and the posterior
standard deviation $\operatorname{Std}[\kappab\vert\yb^\star]$.
\subsection{Additional example: Log-Gaussian Cox process with sparse observations}\label{app:examples:log-cox}

We consider an inference problem in
spatial statistics for a log-Gaussian Cox point process on a square domain 
$\mathcal{D} = [0,1]^2$.
This type of stochastic process
is frequently used to model spatially aggregated point patterns 
\cite{moller1998,Christensen2005,rue2009approximate,girolami2011riemann}.
Following a configuration similar to \cite{Christensen2005,moller1998}, we
discretize $\mathcal{D}$ into a $64 \times 64$ uniform grid, and denote by   
$\sbb_i \in \mathcal{D}$  the center of the
$i$th cell, for $i=1,\ldots,d$,  with $d=64^2$.
We consider a discrete stochastic process $(\Yb_i)_{i=1}^d$, where $\Yb_i$ 
denotes the number of occurrences/points in the $i$th cell.
Each $\Yb_i$ is modeled as a Poisson random variable with mean 
$\exp( \Zb_i )/d$, where $(\Zb_i)$ is a Gaussian process
with covariance
$\cov(\Zb_i,\Zb_j) = 
\sigma^2 \exp \left( - \left\Vert \sbb_i - \sbb_j \right\Vert_2 /
(64 \beta) \right)$
and mean $\Ex[\Zb_i] = \mu$, for all $i=1,\ldots,d$.
We consider the following values for the parameters:
$\beta=1/33$, $\sigma^2 = 1.91$, and $\mu = \log(126) - \sigma^2/2$.
The $(\Yb_i)$ are assumed conditionally independent given the (latent) Gaussian field.
For interpretability reasons, we also 
define the {\it intensity} process $(\Lambdab_i)_{i=1}^d$ as $\Lambdab_i = \exp( \Zb_i )$,
for $i=1,\ldots,d$.
%

\begin{figure}[t]
	\centering
	\begin{subfigure}[b]{.32\linewidth}
		\includegraphics[width=0.9\textwidth]{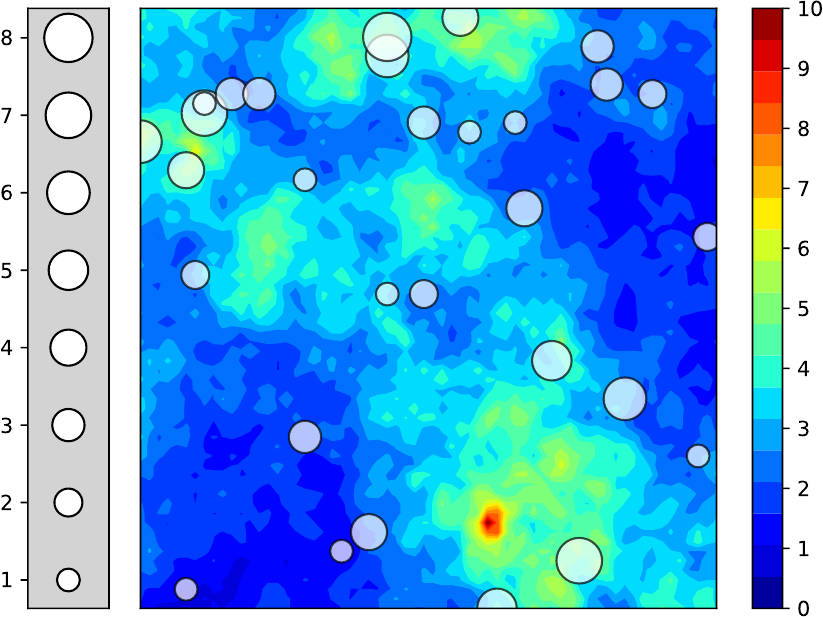}
		\caption{Field $\Lambdab^\star$ and observations $\yb^\star$}
		\label{fig:ex:LogCox:TruthAndRealizations}
	\end{subfigure}
	\hspace{2pt}
	\begin{subfigure}[b]{.28\linewidth}
		\includegraphics[width=0.86\textwidth]{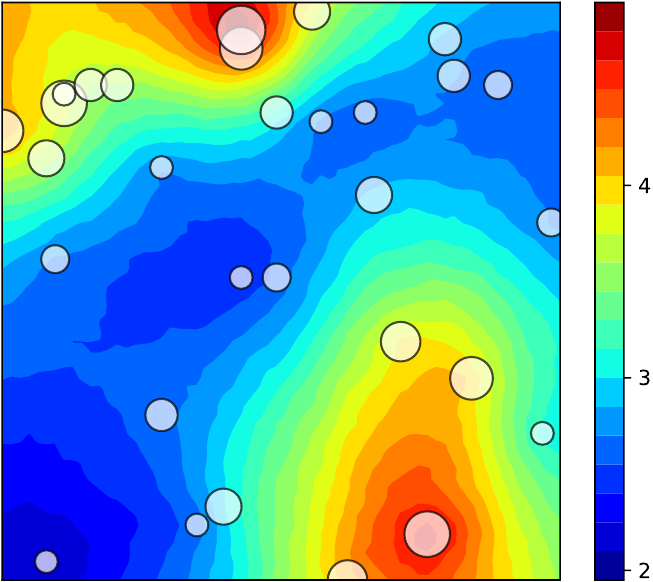}
		\caption{$\mathbb{E}[\Lambdab\vert\yb^\star]$}
		\label{fig:ex:LogCox:posterior-expectation}
	\end{subfigure}
	\hspace{3pt}
	\begin{subfigure}[b]{.32\linewidth}
		\includegraphics[width=.39\textwidth]{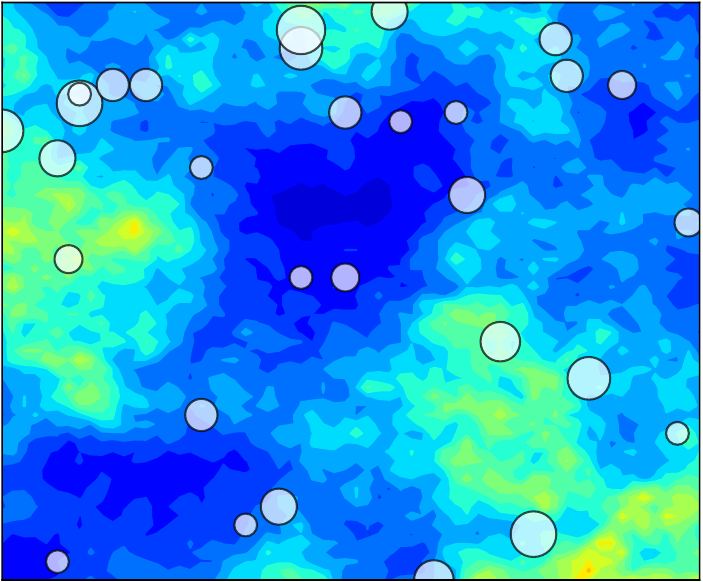}
		\hspace{1pt}
		\includegraphics[width=.39\textwidth]{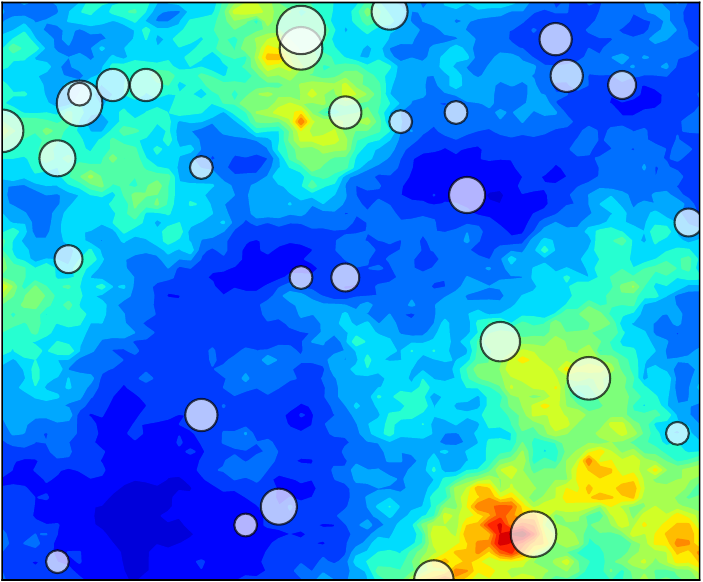}
		\\[2pt]
		\includegraphics[width=.39\textwidth]{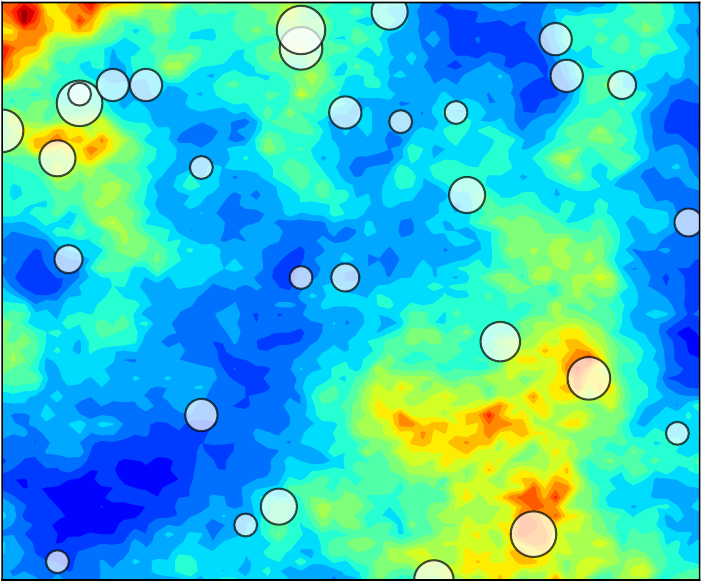}
		\hspace{1pt}
		\includegraphics[width=.39\textwidth]{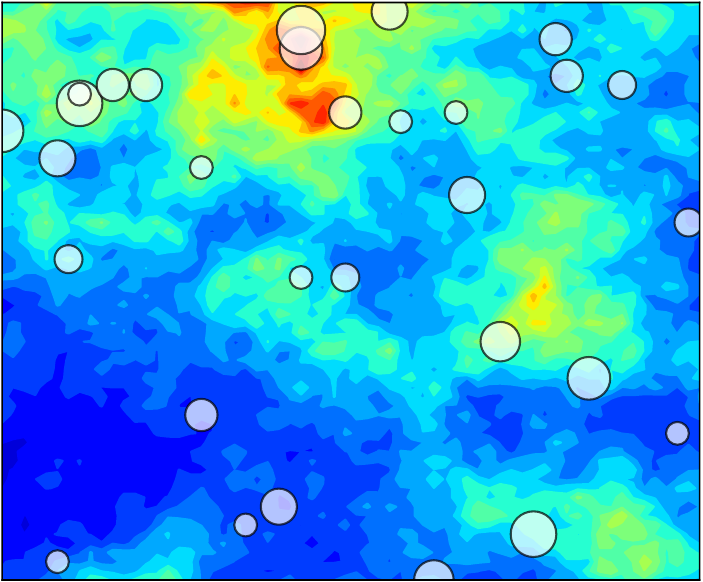}
		\caption{Realizations of $\Lambdab\sim\pi_{\Lambdab\vert\yb^\star}(\lambdab)$}
		\label{fig:ex:LogCox:posterior-realizations}
	\end{subfigure}
	\\[10pt]
	\begin{subfigure}[b]{.32\linewidth}
		\includegraphics[width=\textwidth]{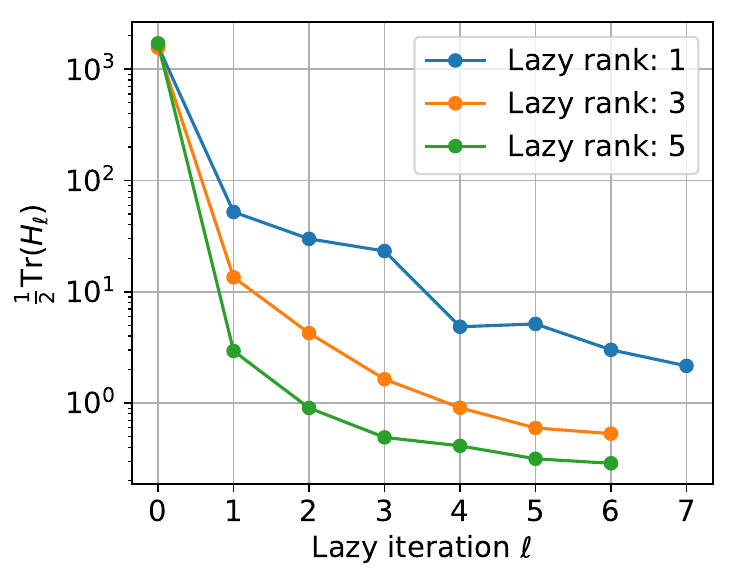}
		\caption{Convergence rate}
		\label{fig:ex:LogCox:convergence}
	\end{subfigure}
	\begin{subfigure}[b]{.32\linewidth}
		\includegraphics[width=\textwidth]{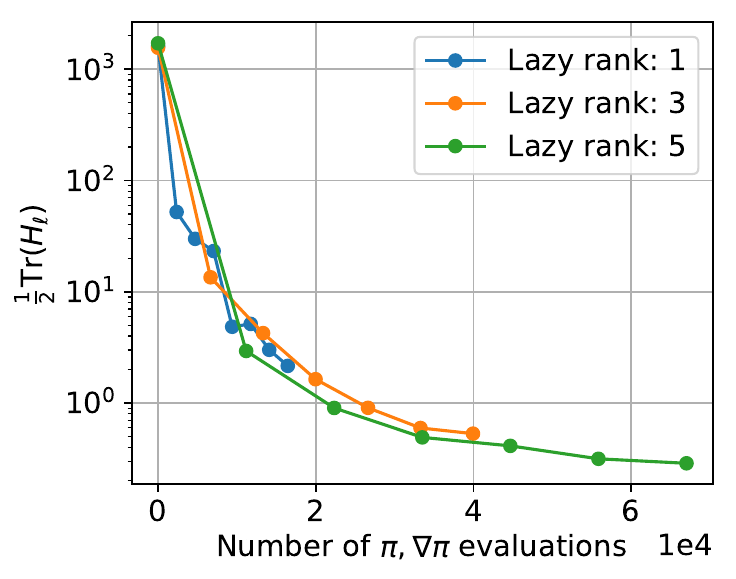}
		\caption{Convergence and cost vs. ranks}
		\label{fig:ex:LogCox:convergence-cost}
	\end{subfigure}
	\hspace{2pt}
	\begin{subfigure}[b]{.31\linewidth}
		\includegraphics[width=\textwidth]{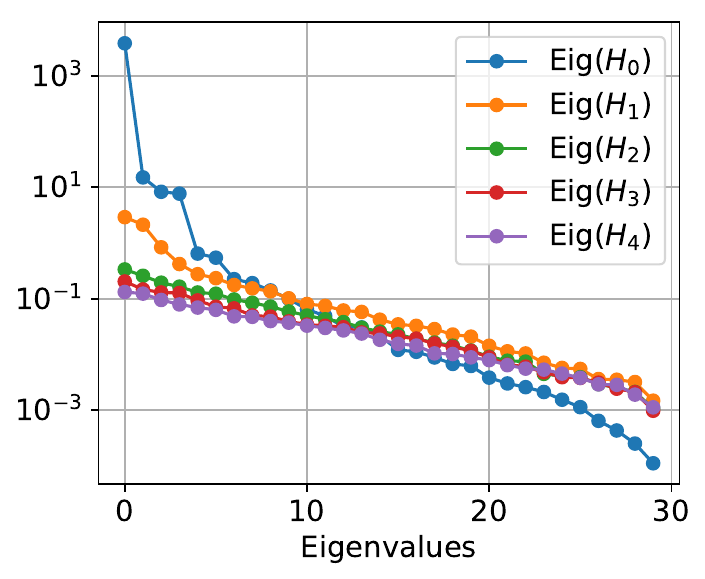}
		\caption{Spectrum decay}
		\label{fig:ex:LogCox:spectrum}
	\end{subfigure}
	
	
	\caption{
		Application of the algorithm on the log-Gaussian Cox
		process distribution.
		Figure (\subref{fig:ex:LogCox:TruthAndRealizations}) shows the intensity field
		$\Lambdab^\star$ used to generate the data $\yb^\star$ (circles).
		Figures
		(\subref{fig:ex:LogCox:posterior-expectation}) shows the
		posterior expectation.
		Figure (\subref{fig:ex:LogCox:posterior-realizations}) shows four
		realizations from the posterior $\pi(\Lambdab\vert\yb^\star)$.
		Figure (\subref{fig:ex:LogCox:convergence}) shows the convergence
		rate of the algorithm as a function of the iterations.
		Figure (\subref{fig:ex:LogCox:convergence-cost}) shows the cost of the
		algorithm for different truncation ranks.
		Figure (\subref{fig:ex:LogCox:spectrum}) shows the decay of the
		spectrum of $H_\ell$ for lazy maps with rank $5$.
	}
	\label{fig:ex:LogCox}
\end{figure}

The goal of this problem is to infer the posterior distribution of the
latent process $\Lambdab \coloneqq (\Lambdab_1,\ldots,\Lambdab_n)$
given 
few sparse realizations of
$\Yb\coloneqq(\Yb_i)$ at $n=30$ spatial locations  $\sbb_{k_1},\ldots, \sbb_{k_n}$ shown in
Figure \ref{fig:ex:LogCox:TruthAndRealizations}.
We denote by $\yb^\star \in \re^n$ a realization of $\Yb$ obtained by sampling
the latent Gaussian field according to its marginal distribution.
Our target distribution is then:
$\pi_{\Lambdab \vert \Yb}({\bm\lambda} \vert \yb^\star)$.

Since the posterior is nearly Gaussian,
we will run three experiements with linear lazy maps and
ranks $r=1,3,5$.
For the three experiments,
the KL-divergence minimized for each lazy
layer and the estimators of $H^\text{B}_\ell$
are discretized with
$m=100,300,500$ Monte Carlo samples respectively.

Figures \ref{fig:ex:LogCox:posterior-expectation}--\subref{fig:ex:LogCox:posterior-realizations}
show the expectation and few realizations of the posterior,
confirming the data provides some
valuable information to recover the field $\Lambdab$.
Figures
\ref{fig:ex:LogCox:convergence}--\subref{fig:ex:LogCox:convergence-cost}
show the convergence rate and the cost of the
algorithm as new layers of lazy maps are added to $\frak{T}_\ell$.
As we expect, the use of maps with higher ranks leads to faster
convergence. On the other hand the computational cost per step
increases---also due to the fact that we increase the
sample size $m$ as the rank increases.
Figure \ref{fig:ex:LogCox:spectrum} reveals the spirit of the
algorithm: each lazy map trims away power from the top of the
spectrum of $H$, which slowly flattens \emph{and} decreases.
To additionally confirm the quality of $\frak{T}_6$ for lazy maps with
rank $5$, and to produce asymptotically unbiased samples from $\pi$, we sample the pullback distribution $\frak{T}_6^\sharp \pi$
using an MCMC chain of length $10^4$, with a Metropolis independence
sampler employing a $\mathcal{N}(0,I_d)$  proposal (see [C. Robert and G. Casella, \emph{Monte Carlo statistical methods}, 2013] for more details).
As explained in \cite{parno2014transport}, the Metropolis independence
sampler is effective insofar as the pullback distribution has
been Gaussianized by the map.
The reported
acceptance rate is $72.6\%$ with the worst effective sample size
(over all $d=4096$ chain components) being
$26.6\%$ of the total chain.

\subsection{Additional example: Estimation of the Young's modulus of a cantilever beam}
\label{app:examples:timoshenko}

\begin{figure}
	\centering
	\begin{subfigure}[b]{.31\linewidth}
		\includegraphics[width=\textwidth]{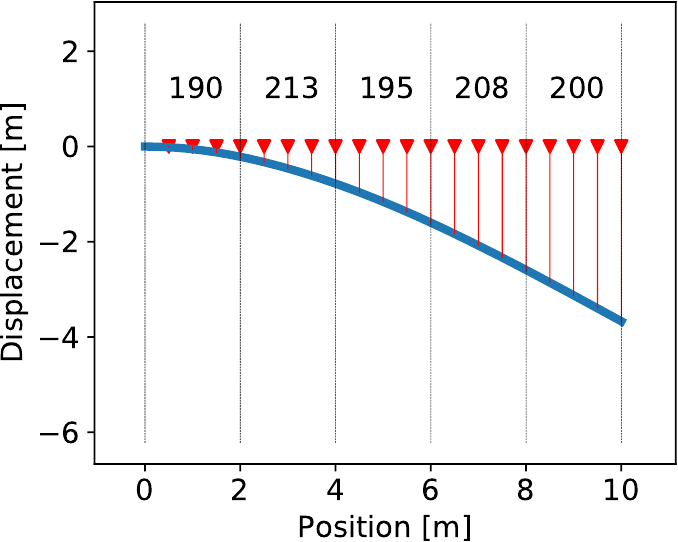}
		\caption{Experimental setting}
		\label{fig:ex:cantilever:experiment-setting}
	\end{subfigure}
	\hspace{3pt}
	\begin{subfigure}[b]{.31\linewidth}
		\includegraphics[width=\textwidth]{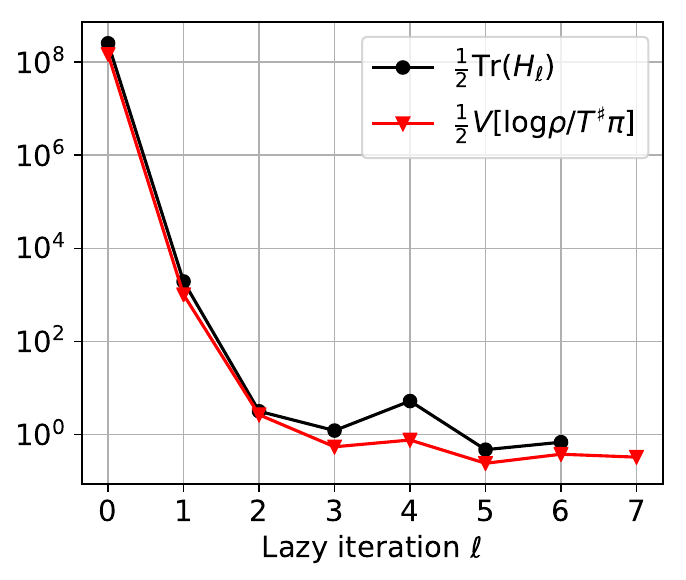}
		\caption{Convergence rate}
		\label{fig:ex:cantilever:convergence}
	\end{subfigure}
	\hspace{5pt}
	\begin{subfigure}[b]{.28\linewidth}
		\includegraphics[width=\textwidth]{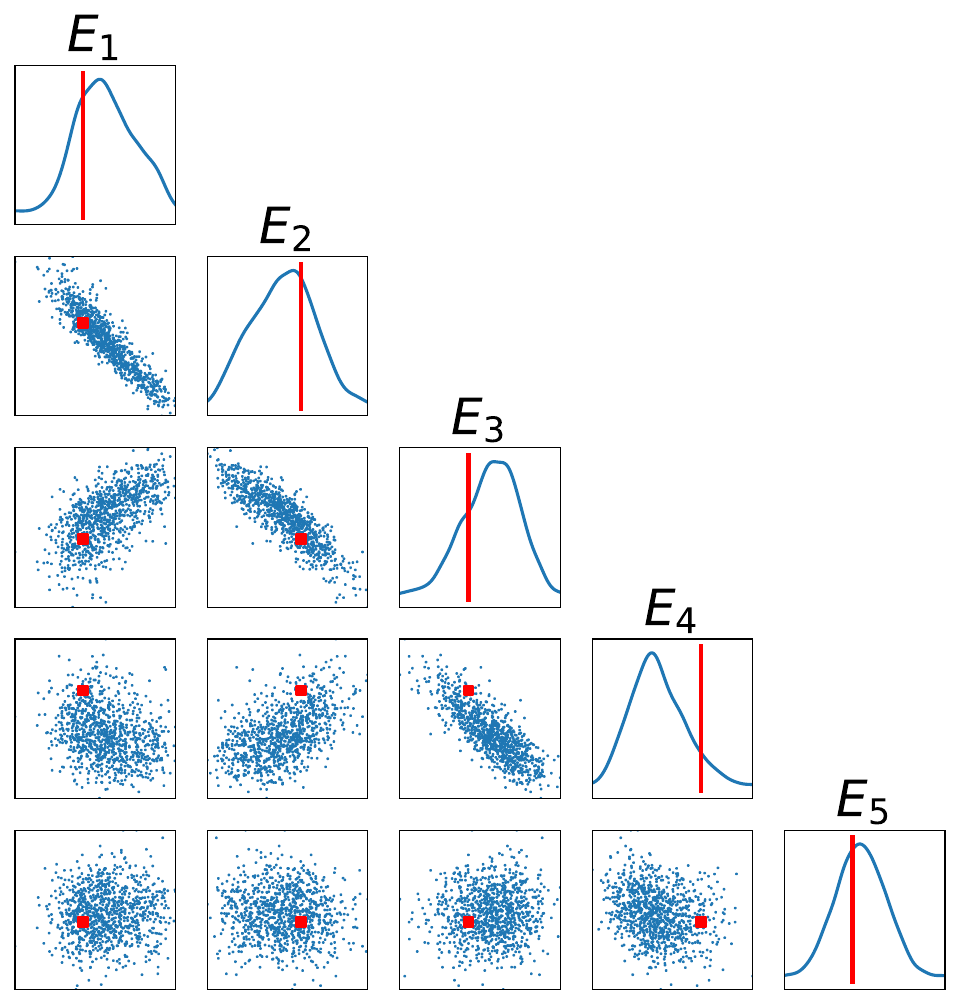}
		\caption{Marginals of $\pi(\Eb\vert\yb^\star)$}
		\label{fig:ex:cantilever:aligned-marginals}
	\end{subfigure}
	\caption{
		Application of the algorithm for the estimation of the Young's
		modulus of a cantilever beam.
		Figure (\subref{fig:ex:cantilever:experiment-setting}) shows the
		experimental setting with the beam clamped at $x=0$, the load
		applied at $x=l$, $20$ sensors marked in red, and the true Young's
		modulus [$\SI{}{\giga\pascal}$] for each segment.
		Figure (\subref{fig:ex:cantilever:convergence}) shows the
		convergence of the algorithm in terms of the variance and trace diagnostics.
		Figure (\subref{fig:ex:cantilever:aligned-marginals}) shows
		marginals of the posterior distribution $\pi(\Eb\vert\yb^\star)$
		along with the true values (red).
	}
	\label{fig:ex:cantilever}
\end{figure}

Here we consider the problem of estimating the Young's modulus $E(x)$
of an inhomogeneous cantilever beam, i.e., a beam clamped on one side ($x=0$) and
free on the other ($x=l$).
The beam has a length of $l=\SI{10}{\meter}$,
a width of $w=\SI{10}{\centi\meter}$ and
a thickness of $h=\SI{30}{\centi\meter}$.
Using Timoshenko's beam theory, the displacement $u(x)$
of the beam under the load $q(x)$ is modeled by the coupled PDEs
\begin{equation}
\label{eq:ex:timoshenko-pde}
\begin{cases}
\ddx \left[ \frac{E(x)}{2(1+\nu)} \left( \varphi(x) - \ddx w(x) \right) \right] = \frac{q(x)}{\kappa A} \;, \\
\ddx \left( E(x) I \ddx \varphi(x) \right) = \kappa A \frac{E(x)}{2(1+\nu)} \left( \varphi(x) - \ddx w(x) \right) \;,
\end{cases}
\end{equation}
where  $\nu=0.28$ is the Poisson ratio,
$\kappa=5/6$ is the Timoshenko shear coefficient for rectangular
sections, $A=wh$ is the cross-sectional area of the beam, and
$I=wh^3/12$ is its second moment of inertia. We consider a beam composed of $d=5$ segments each of $\SI{2}{\meter}$ length
made of different kinds of steel,
with Young's moduli
$\Eb^\star=\{E_i\}_{i=1}^5 = \{190, 213, 195, 208, \SI{200}{\giga\pascal}\}$ respectively,
and we run the virtual experiment of applying
a point mass of $\SI{5}{\kilo\gram}$ at the tip of the beam.
Observations $\yb^\star$ of the displacement $w$ are gathered at the
locations shown in Figure
\ref{fig:ex:cantilever:experiment-setting} with a measurement noise of $\SI{1}{\milli\meter}$. 
We endow $\Eb$ with the prior
$\pi(\Eb)=\Nc(\SI{200}{\giga\pascal},25\cdot I_5)$ and
our goal is to characterize the posterior distribution
$\pi(\Eb\vert\yb^\star) \propto \Lc_{\yb^\star}(\Eb)\pi(\Eb)$. Let $\Sc$ be the map $\Eb \mapsto w$ delivering the solution to (\ref{eq:ex:timoshenko-pde}).
Observations are gathered through the operator $B_i(w)\coloneqq\int_0^l w\,\phi_i\,\d\xb$,
where $\phi_i$ are defined the same way as in Appendix \ref{app:examples:elliptic}
for locations $\{s_i \coloneqq i \cdot 0.5\}_{i=1}^{20}$.
Defining the parameter-to-observable map $\Fc:\Eb\mapsto [B_1(\Sc(\Eb)),\ldots,B_{20}(\Sc(\Eb))]^\top$,
observations $\yb$ are assumed to satisfy the model $\yb = \Fc(\Eb) + \epsilon$,
where $\epsilon \sim \Nc(0, 10^{-6} \cdot I_{20})$ corresponds to $\SI{1}{\milli\meter}$
of measurement noise.

\begin{figure}
	\centering
	\begin{subfigure}[b]{.31\linewidth}
		\includegraphics[width=\textwidth]{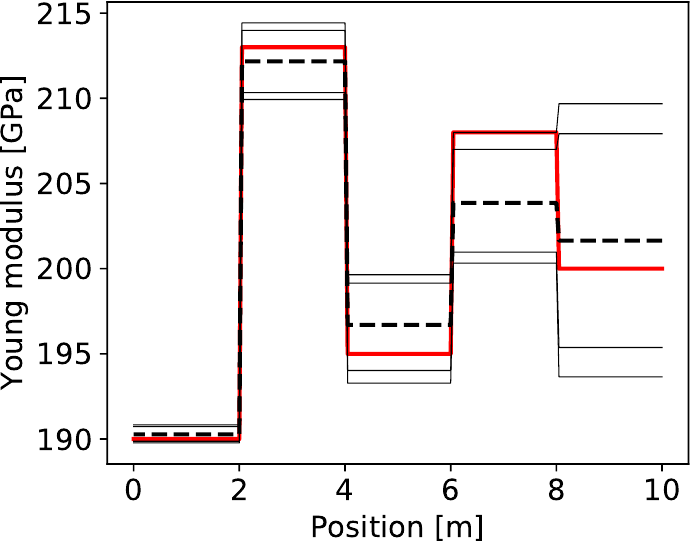}
		\caption{True field vs. posterior}
		\label{fig:ex:cantilever:true-vs-posterior}
	\end{subfigure}
	\hspace{3pt}
	\begin{subfigure}[b]{.31\linewidth}
		\includegraphics[width=\textwidth]{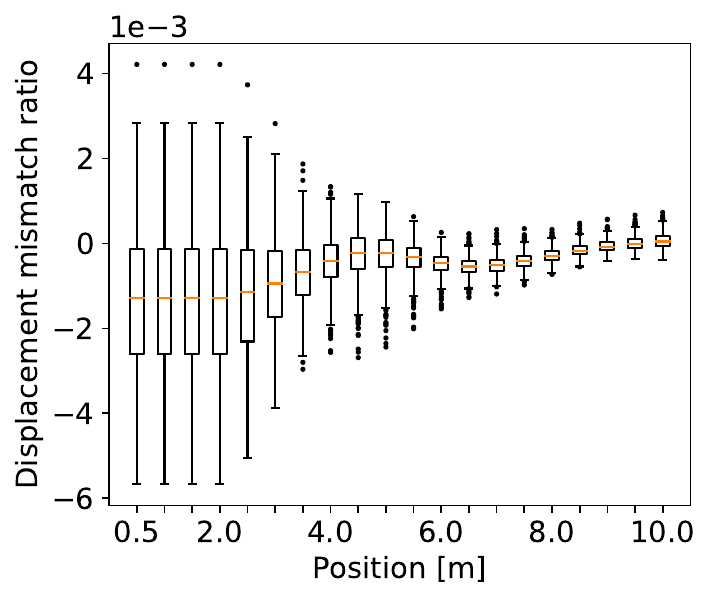}
		\caption{Posterior predictive mismatch}
		\label{fig:ex:cantilever:post-pred}
	\end{subfigure}
	\caption{
		Additional results for the estimation of Young's
		modulus of a cantilever beam.
		Figure (\subref{fig:ex:cantilever:true-vs-posterior}) shows the
		mean (dashed black) and the $5,10,90,95$-percentiles (thin black)
		of the marginals of $\pi(\Eb\vert\yb^\star)$ compared with the true values
		(red).
		Figure (\subref{fig:ex:cantilever:post-pred}) shows the
		distribution of $(\yb^\star - \yb)/\vert\yb^\star\vert$,
		where $\yb$ is distributed
		according to the posterior predictive
		$\pi(\yb\vert\yb^\star)=\pi(\yb\vert \Eb)\pi(\Eb\vert\yb^\star)$.
	}
	\label{fig:app:ex:cantilever}
\end{figure}

The algorithm is run with rank $2$ lazy maps using triangular
polynomial maps  of degree $3$ as the underlying transport class. The
expectations appearing in the algorithms are approximated using $m=100$ samples from $\Nc(0,I_5)$.
Figures \ref{fig:ex:cantilever} and \ref{fig:app:ex:cantilever} summarize the results.
We further confirm these results by generating an MCMC chain of length
$10^4$ using Metropolis-Hastings with a $\mathcal{N}(0,I_d)$
independence proposal; the target distribution for MCMC is the
pullback $\frak{T}_\ell^\sharp\pi$, as in previous examples.
The reported acceptance rate is $68.3\%$
with the worst, best, and
average effective sample sizes being $7.0\%$, $38.7\%$, and $20.1\%$
of the complete chain. 
In this example we fix
the Poisson ratio, but one could think of it varying from material to
material, and thus estimate it jointly with the Young's modulus.

\end{document}